\documentstyle[11pt,epsfig]{article}

\newcommand{\bra}[1]{\langle #1 \mid}
\newcommand{\ket}[1]{\mid #1 \rangle}
\newcommand{\braket}[2]{\langle #1 \mid #2 \rangle}

\newcommand{\f}{\begin{equation}}
\newcommand{\ff}{\end{equation}}
\def \ut#1{\rlap{\lower1ex\hbox{$\sim$}}#1{}}

\newcommand{\pic}[5]{\raisebox{#3pt}{
\hspace{#4pt}\psfig{file=#1.ps,height=#2pt,silent=}\hspace{#5pt}}}

\newcommand{\kd}[1]{\mathchoice{
\pic{#1}{35}{-15}{0}{0}}{
\pic{#1}{12}{-2}{-3}{2}}{
\pic{#1}{9}{-2}{-3}{1}}{
\pic{#1}{7}{-1}{-1}{0}}}
\begin{document}

\title{The geometry of quantum spin networks}

\author{ \\ \\
\small {\bf Roumen Borissov${}^{*,1,2}$}, {\bf Seth Major${}^{\#,1}$ },
and
{\bf Lee Smolin${}^{\dagger,1}$} \\ \\
\small ${}^1$ Center for Gravitational Physics and Geometry,
The Pennsylvania State University\\
\small University Park, PA 16802 \\ \\
\small ${}^2$ Department of Physics, Temple University, \\
\small Philadelphia, PA 19122. }
\date{}
\maketitle

\vspace{0.5cm}

\begin{abstract}
The discrete picture of geometry arising from the
loop representation of quantum gravity can be extended by a
quantum deformation. The
operators for area and volume defined in the q-deformation of the theory are
partly diagonalized. The eigenstates are expressed in terms of $q$-deformed
spin networks. The $q$-deformation breaks some of the
degeneracy of the volume operator so that trivalent spin-networks
have non-zero volume. These computations are
facilitated by use of a technique based on the recoupling theory of
$SU(2)_q$, which simplifies the construction of these and
other operators through diffeomorphism invariant regularization
procedures.
\end{abstract}

\vskip .3cm
\noindent
\vfill
$\\{}^*$ borissov@astro.ocis.temple.edu,
$\\{}^\#$seth@phys.psu.edu,
$\\{}^\dagger$  smolin@phys.psu.edu.

\section{Introduction.}

Topological quantum field theory arose from a study of
invariants of three and four dimensional
manifolds (\cite{witten-tqft} - \cite{rogerb}). This
led to a new realm of mathematical structure,
which ties together topology, representation theory and
category theory.
In three
dimensions, the Chern-Simons topological quantum field theory
is closely related to invariants of knots, links and networks. Three
dimensional
topological quantum field theories also
provide models of quantum
gravity in $2+1$ dimensions which teach us about the
construction and interpretation of diffeomorphism invariant
quantum field theories \cite{2+1}. However, from the beginning, there have
been reasons to believe that three dimensional topological
quantum
field theories might play a direct role in quantum gravity in $3+1$
dimensions. One reason for this is that the
Kodama state \cite{kodama}, which is the exponential of the Chern-Simons
invariant,
is the only state known to be an exact physical state
of quantum gravity and, in the same time, to have a semi-classical
interpretation
as the vacuum state associated to  DeSitter spacetime\cite{lscs}.
Another reason is that in the presence of a boundary, conditions may
be chosen so that Chern-Simon theory is induced from quantum
gravity in
the same way that, one dimension lower, Wess-Zumino-Witten
theory is
induced on the boundary of
Chern-Simons theory \cite{linking,lsbekenstein}. Still other
arguments for a role for Chern-Simons theory in quantum gravity
come from the interpretation problem in quantum
cosmology\cite{louis-int}, from the holographic hypothesis\cite{holographic}
and from the attempts to construct
quantum gravity algebraically as an extension (or categorification)
of it\cite{louisigor,baezdolan, louis-int}\footnote{We may
also mention an interesting paper
in which it seems possible also to derive four dimensional quantum
gravity as the boundary theory of a five dimensional TQFT\cite{rogerb}.}.

Recently another reason for suspecting a close relation between
quantum gravity and topological quantum field theory has emerged. Spin
networks  play a key role in the states in both formalisms.
(A spin network is a closed graph with
edges labeled by the representations of $SU(2)$ and
vertices labeled by the ways the representations joined at the vertex can be
combined into a singlet \cite{roger-sn}.) In canonical quantum gravity
(\cite{A1} - \cite{ls-review}) it has been
discovered that there is a basis of spatially diffeomorphism invariant
states of the gravitational field which are labeled by spin
networks (\cite{RS2} - \cite{spinnet-john}).\footnote{Note that
the elements of the basis
must be differentiated by labels attached to vertices of valence higher
than three to resolve the degeneracy.}
This construction proved to be an important step
in the search for physical states. It is also  an
essential ingredient in the development of the measure theory
on the space of connections modulo gauge transformations
\cite{ALTMM}. A closely related structure, quantum
spin networks, play a basic role in topological quantum
field theory \cite{lou-qnet,KL}. In a quantum spin network
the edges are labled by
representations of some quantum group and the vertices
are labeled by the corresponding intertwiners. It is possible that
spin networks provide a bridge from quantum
gravity to topological quantum field theory and conformal field
theory, and perhaps even to string theory.  But if this is to happen,
quantum gravity
itself must be expressed in terms of quantum spin networks.  It
turns out that this can be done directly \cite{MS}.
In this alternative quantization, framing factors arise at
 order $\hbar$
that modify the naive algebra of products of holonomies. One result
is that one can no longer multiply operators
associated to Wilson loops at will, as one can in the classical theory.
For example, there is an $n$ such that for any loop $\alpha$,
$\left(\hat{T}[\alpha ]\right)^n =0$. It is expected that this alternate
quantization is appropriate to the case of non-vanishing cosmological
constant, $\Lambda$, in which case the deformation parameter
$q$ of the quantum group is given by \cite{MS}
\f
q= e^{ i \pi  \over r}  = e^{i \hbar^2 G^2 \Lambda /6}.
\ff

More generally, we may note that any quantization of a non-canonical
algebra such as the loop algebra involves a deformation of the
classical algebra by factors proportional to $\hbar$. When we consider the
``higher" loop operators, with more than one insertions of the frame
field $\tilde{E}^{ai}$ then the quantum algebra is necessarily
deformed by such factors \cite{RS1}. The alternative quantization involves
simultaneously deforming in the factor $q$ and $\hbar$ providing an alternative
to the
usual quantization such that the quantum observables' algebra reduces to
the classical one as $\hbar \rightarrow 0$.

Using this alternative quantization, we may express physically
interesting observables directly in terms of operators in this
q-deformed quantum theory. Among these are kinematical observables
such as volume and area. If we label these regions and
surfaces by the values of physical fields these may be promoted to
diffeomorphism invariant observables.

The primary purpose of this work is to construct several observables
for the alternate, q-deformed quantization. The result may be
considered to endow quantum spin networks with a new physical interpretation
in terms of three dimensional quantum geometry. For given any quantum
spin network we are able to associate areas to its edges and volumes to its
vertices.
In related papers, these results will be extended to dynamical
operators such as the hamiltonian and hamiltonian constraint of quantum
gravity \cite{RB,BRS}.

We may note several advantages of the formulation we present here.
First, we use techniques from the Temperly-Lieb
recoupling theory \cite{KL}.  These have strong practical
advantages as they provide an elegant and efficient way to do
calculations involved in the construction and action of operators in
non-perturbative quantum gravity. Second, some problems associated
with ordinary spin network states
in non-perturbative quantum gravity seem to be ameliorated by the
quantum deformation. In particular, the volume operator has a great
deal of degeneracy in the ordinary representation, in that trivalent
vertices all contribute zero volume \cite{RL}. We find that the quantum
deformation lifts this degeneracy. This may make possible several
developments which we mention in the conclusion.

The next three sections summarize the construction of the
space of states in q-deformed quantum gravity as well
as the basic hypothesis that underlies the regularization
procedure in this theory \cite{MS}. This is followed by a short
summary of the recoupling theory of quantum spin networks
from \cite{KL}. After these preliminaries, we show in Sections
6 to 8 how to define, regulate and compute the area and volume
operators in q-deformed quantum gravity. We close with some
comments on the applications of these results and techniques
which are currently under study.

\section{The state space of q-deformed quantum gravity}

We first briefly describe the structure of the space ${\cal H}^q$ of
quantum states of the gravitational field after the $q$-deformation.
More details may be found in \cite{MS}. The space has an orthonormal
basis of states $\ket{\alpha}$ labeled
by distinct quantum spin networks $\alpha$. A quantum spin
network  (or q-spin net) consists of the embedding of closed graph
into the three manifold $\Sigma$ (with fixed topology) with edges
labeled by the representations of $SU(2)_q$ and vertices
labeled by distinct ways to decompose the incoming representations
into a singlet (trivalent vertices are unique and are thus unlabeled.)

The deformation parameter $q=e^{ i \pi /r}$ will be taken to be
at a root of unity, in which case the representations are labeled
by integers, $j$, denoting twice a spin. We will also
find it convenient to parameterize the deformation by $A$ such
that $A^2 =q$. The usual loop representation is then recovered
in the limit $A \rightarrow -1$ (in
the binor convention in which $-1$ has been inserted into each
$SU(2)$ trace). To avoid confusion we will call this the ``ordinary''
case.

The edges of the spin network may be thought of as being
decomposable into framed loops. This is described in \cite{MS},
where it is shown that straight intersections may be decomposed into
linear combinations of two linearly independent intersections,
called the ``over'' and ``under'' touch, and are denoted
$\kd{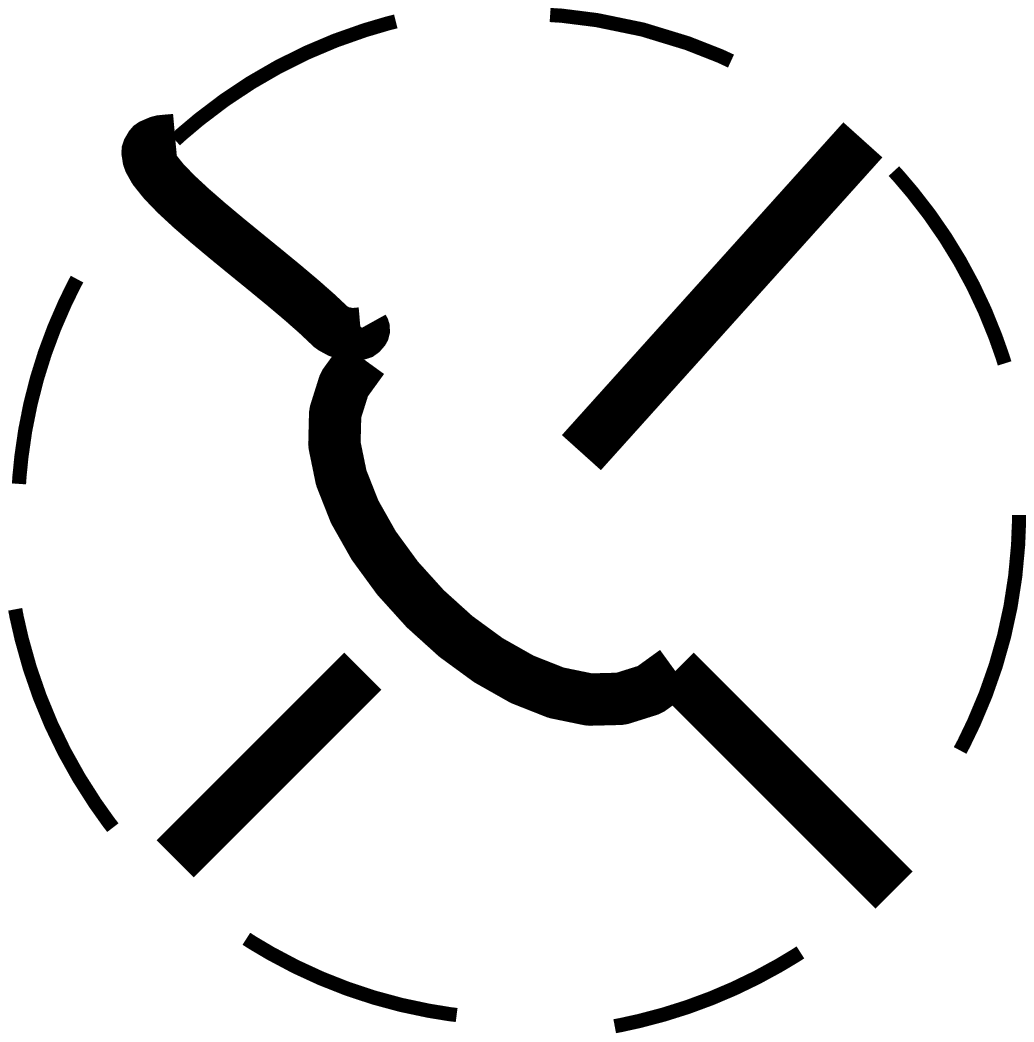}$ and $\kd{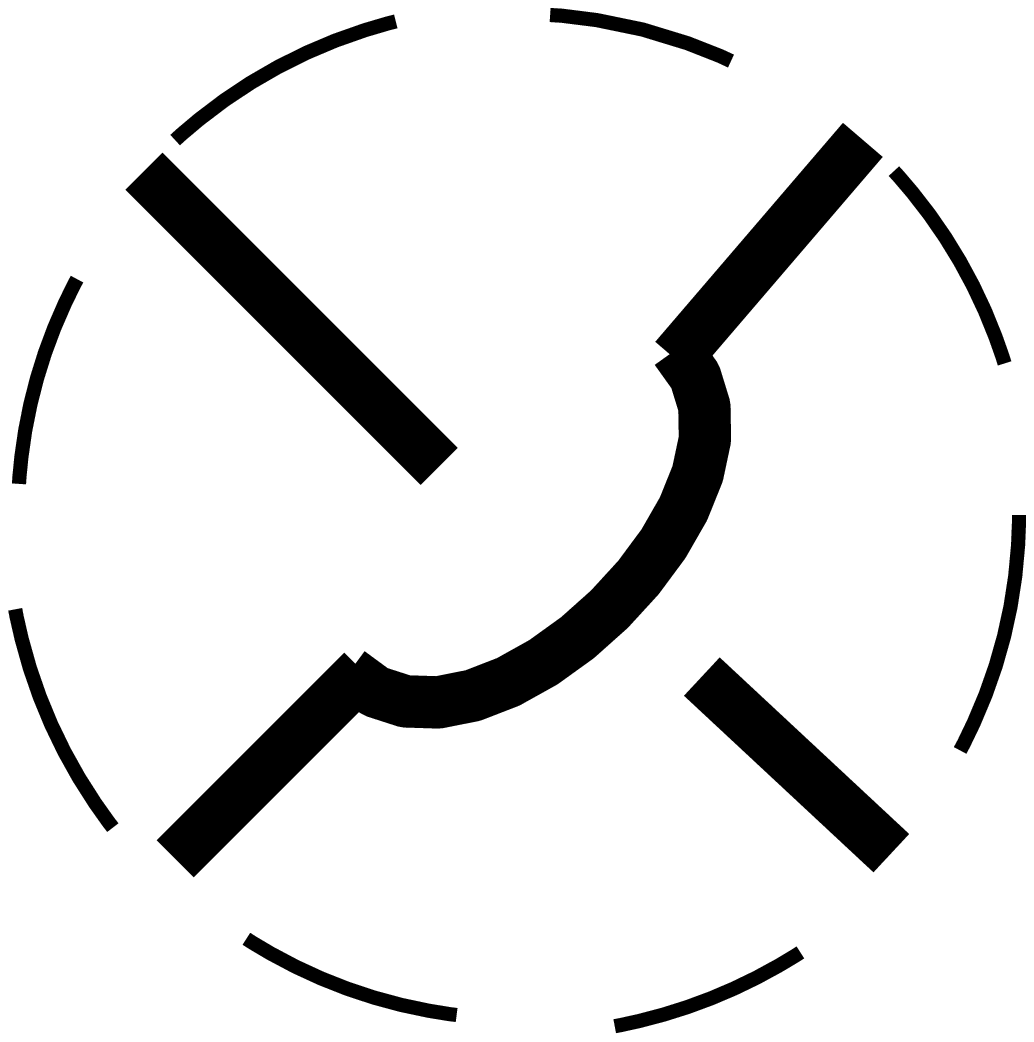}$, respectively. To have consistency
with the Kauffman bracket \cite{LK} the
Mandelstam identity is extended to the two relations
\begin{eqnarray}
\kd{uptouch} &=& A^{-1} \kd{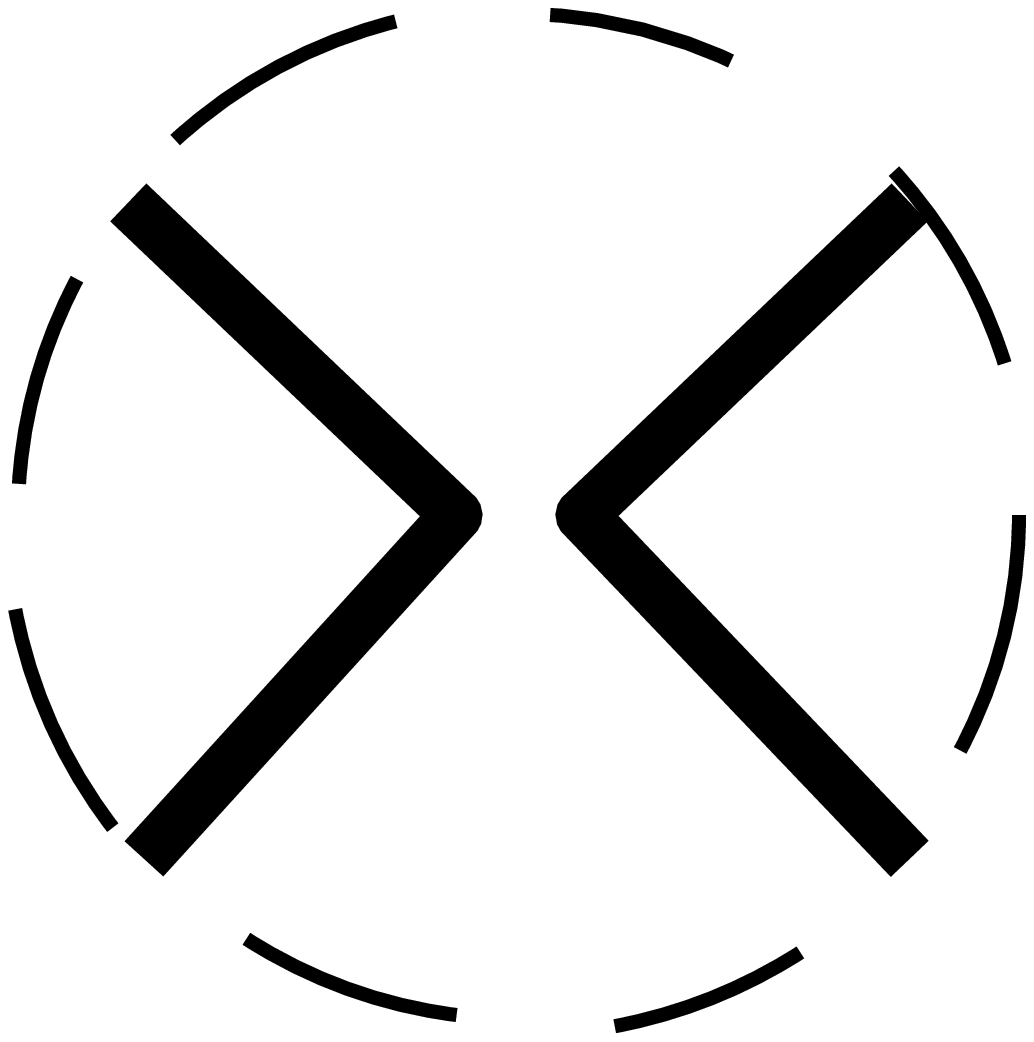} + A \kd{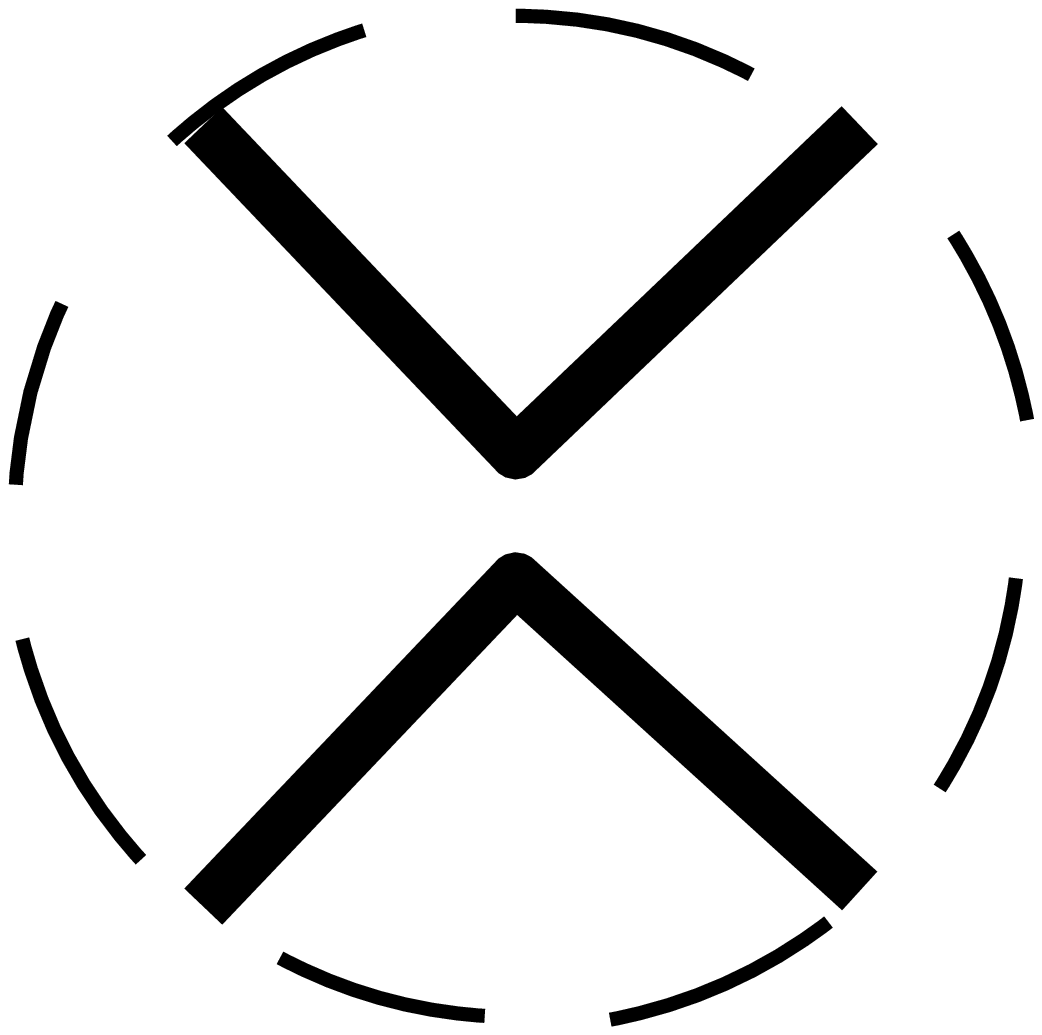}
\label{upmandle}
\\
\kd{dntouch} &=& A \kd{collision} + A^{-1} \kd{wedges}.
\label{dnmandle}
\end{eqnarray}
The resulting structure is a graph with labeled edges represented
diagramically as
\begin{equation}
\pic{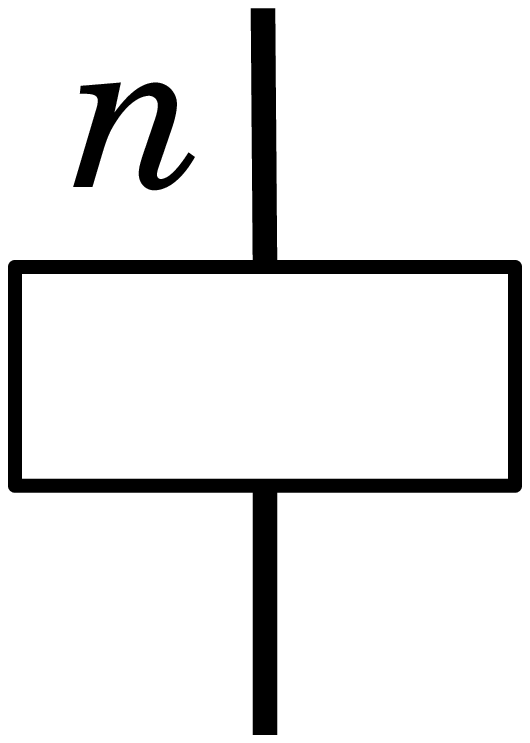}{30}{-15}{-1}{2}= { A^{2n-2} \over [n]!} \sum_{\sigma \in S_n}
\left(A^{-3}
\right)^{t(\sigma)} \pic{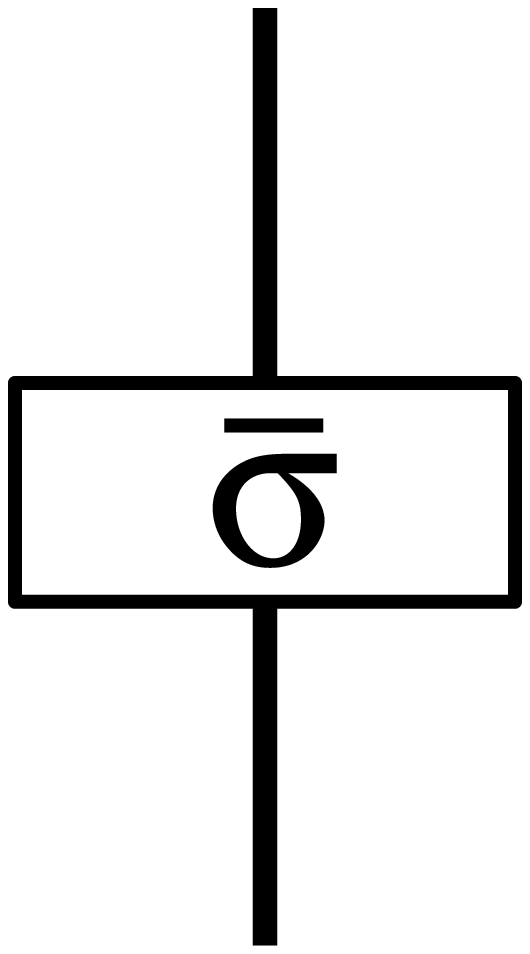}{35}{-15}{-1}{2}
\label{qsym}
\end{equation}
where $\bar{\sigma}$ refers to the right handed braidings \cite{KL} and
the ``quantum integer'' $[ n ]$ is defined by
\f\label{qint}
[n]={q^{n}-q^{-n} \over q-q^{-1}}={A^{2n}-A^{-2n} \over A^{2}-A^{-2}}.
\ff

\begin{figure}
\begin{tabular*}{\textwidth}{c@{\extracolsep{\fill}}c@{\extracolsep{\fill}}
c@{\extracolsep{\fill}}}
\pic{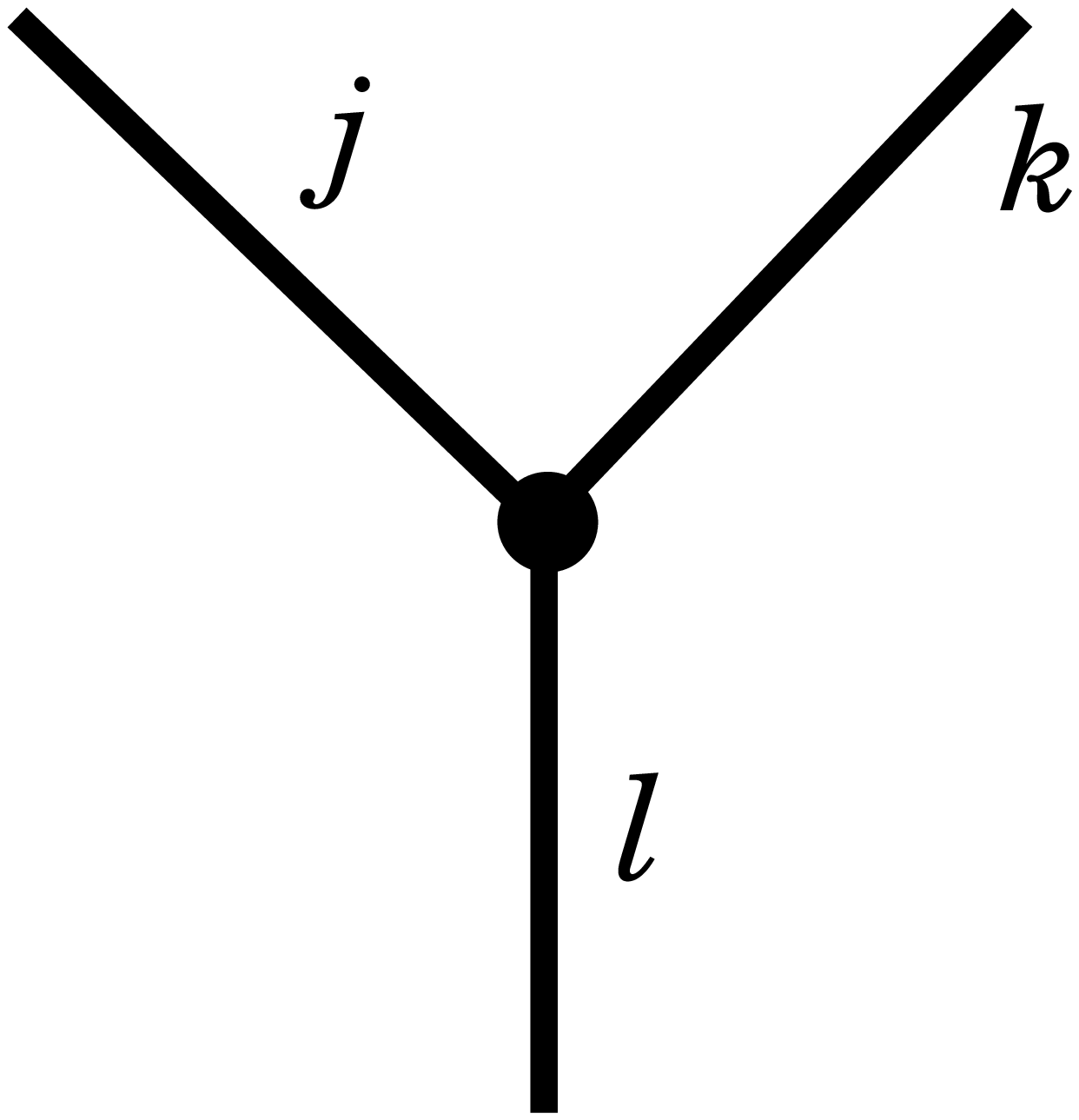}{70}{-5}{20}{0} & \pic{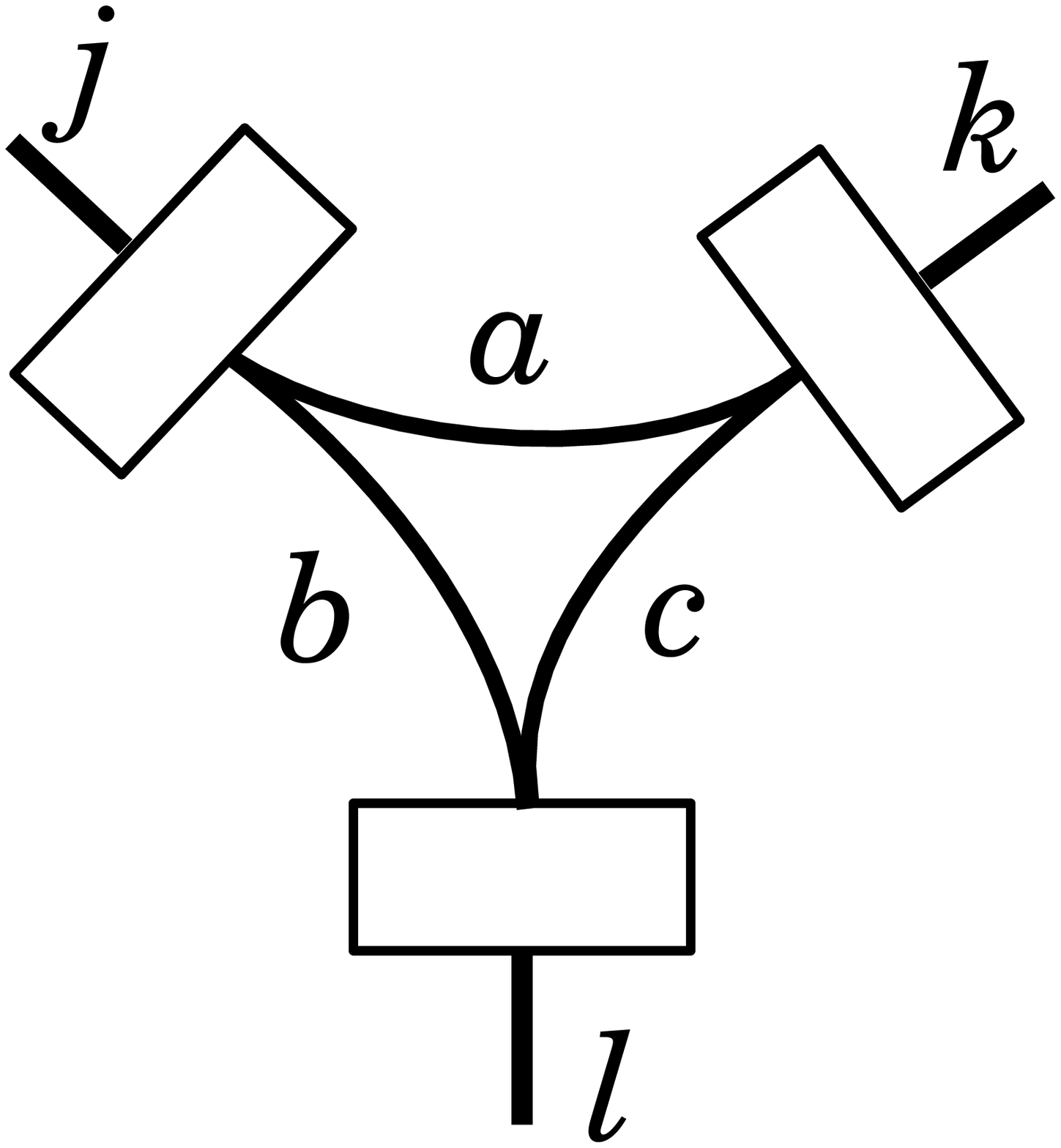}{80}{-5}{0}{20}
& \pic{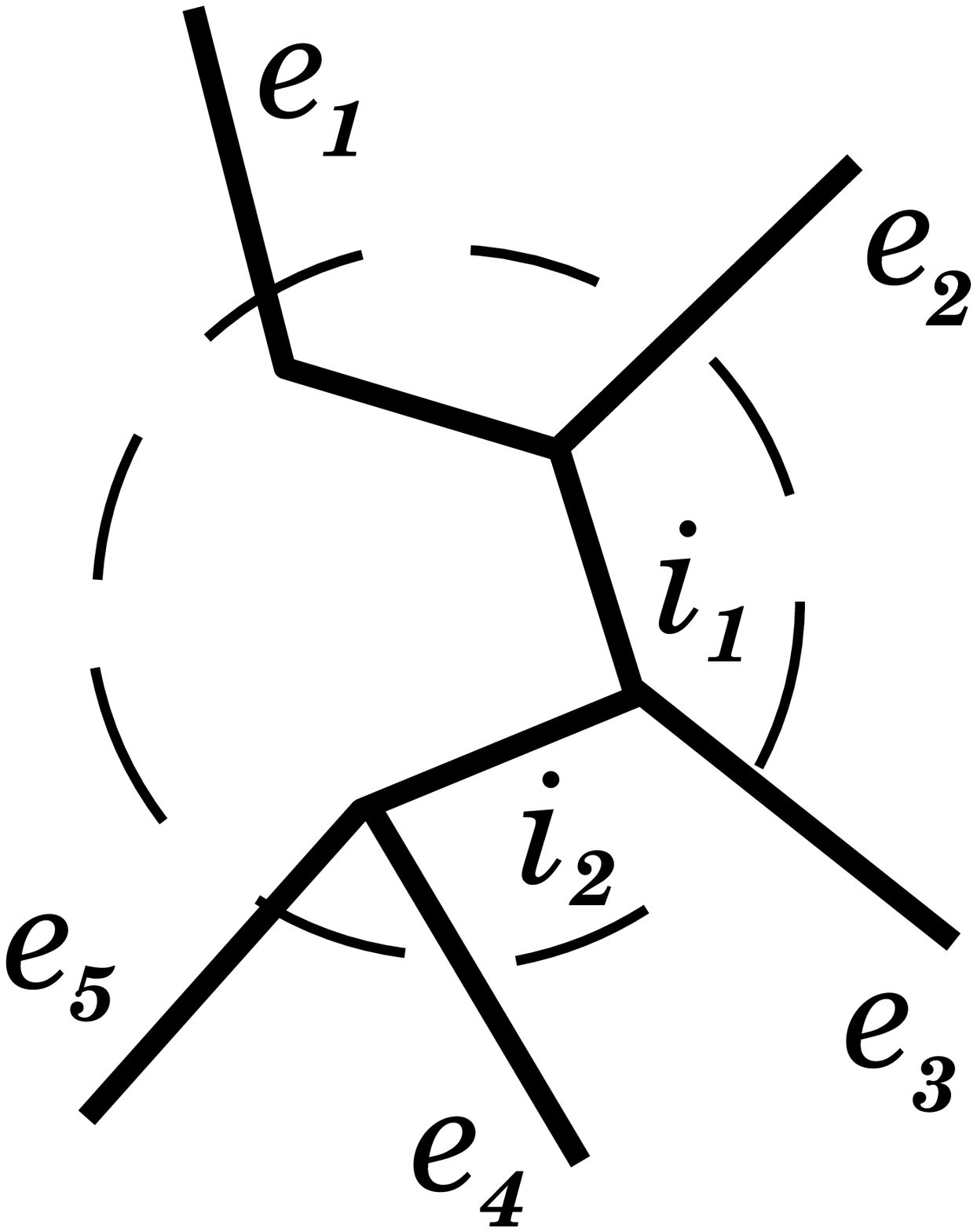}{80}{-15}{20}{0} \\
a & b & c\\
\end{tabular*}
\caption{The trivalent vertex (a.) is decomposed into three projectors
as in (b.) with $a = (j+k-l)/2$, $b=(k+l -j)/2$, and $c=(j+l-k)/2$.  Higher
valent intersections may be decomposed in terms of trivalent ones as in
(c).}
\label{vertex}
\end{figure}

Intertwining operators at vertices of the q-spin net label
the way the different loops pass through the vertex. In the case of
trivalent vertex the intertwining operator is trivial because there is
a unique way for the loops to pass through
the vertex. This can be seen on Fig. (\ref{vertex}b) where, for instance,
$a$ is the number of loops shared by the $j$-edge to the $k$-edge.
In the case of vertices joining more than three edges,
this
decomposition is not available and we need additional information
for the
structure of the vertex. One of the ways to describe such a vertex is
to
first form a set of trivalent vertices joined with ``internal'' lines as
in Fig. (\ref{vertex}c). All these ``internal'' lines have zero length. Then
the vertex is labeled by the color of the ``internal" lines as well as by
topological factors such as those illustrated in Fig. (2) associated
with how the four vertex is defined.

As the $q$-spin networks comprise a linearly independent basis,
they will also label a set of bras, $\bra{\Gamma}$ such that a general
state is given by
\f
\Psi [\Gamma ] = \braket{\Gamma}{\Psi}.
\ff
${\cal H}^q$ may be endowed with an inner product  so that
\f
\braket{\Gamma}{\Gamma^\prime} = \delta_{\Gamma \Gamma^\prime}.
\ff
Following the procedure for ordinary spin network states \cite{RS2},
we may also define a space
${\cal H}_{diffeo}^q \subset {\cal H}^q$ of diffeomorphism invariant
states.  These are labled by an orthonormal basis $\ket{\{\Gamma \}}$
where $\{ ...\}$ stands for diffeomorphism equivalence class defined
by
\f
<\Gamma^\prime |\{\Gamma \}> =
\delta_{\{ \Gamma^\prime \} \{ \Gamma \}}.
\ff

\section{Operators in $q$-deformed quantum gravity}

In the $q$-deformed quantum gravity the basic operators
$\hat{T_q}[\alpha ]$ are associated with framed loops $\alpha^f$.  The
framing is necessary to represent the behavior of Wilson loops
in the presence of the
Chern-Simon measure \cite{MS}.   These act as,
\f
\langle \Gamma |\hat{T}_{q}[\alpha] = \langle \Gamma \cup\alpha |.
\ff
where $\cup $ is a commutative product among quantum spin
networks defined by decomposing using the
edge addition formula \cite{KL},
\begin{equation}
\pic{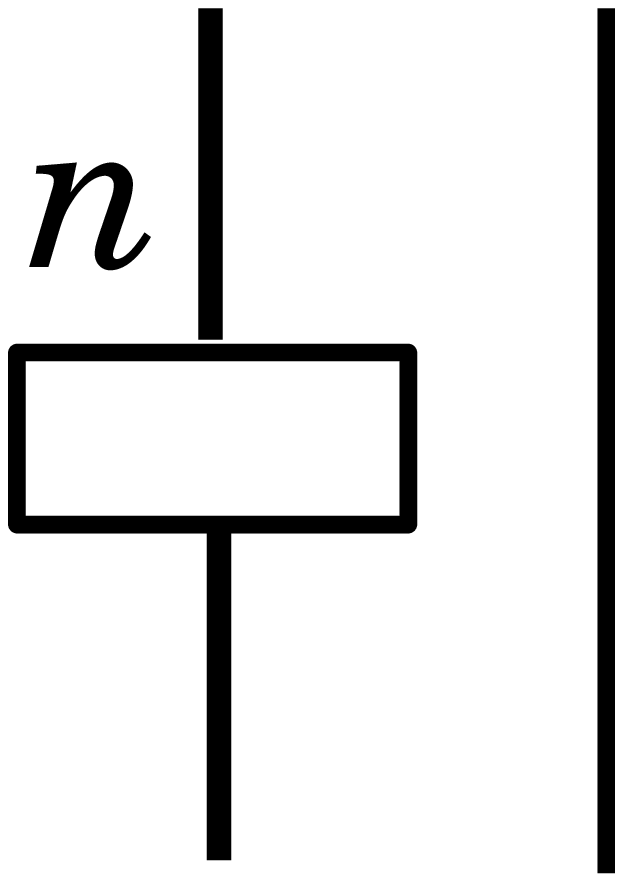}{35}{-15}{-1}{2} =\pic{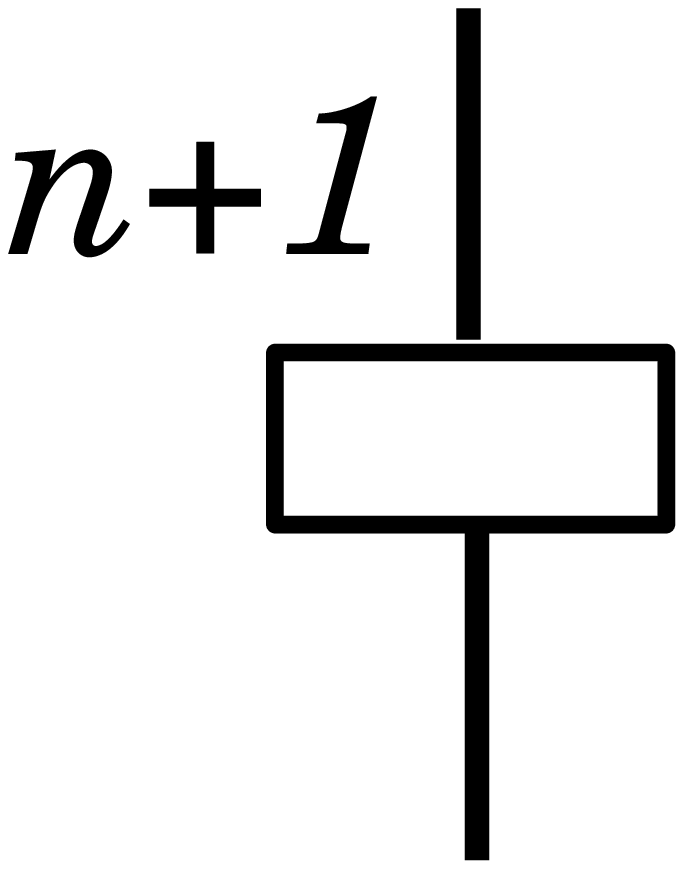}{35}{-15}{-1}{2} -
{ [n] \over [n+1]} \pic{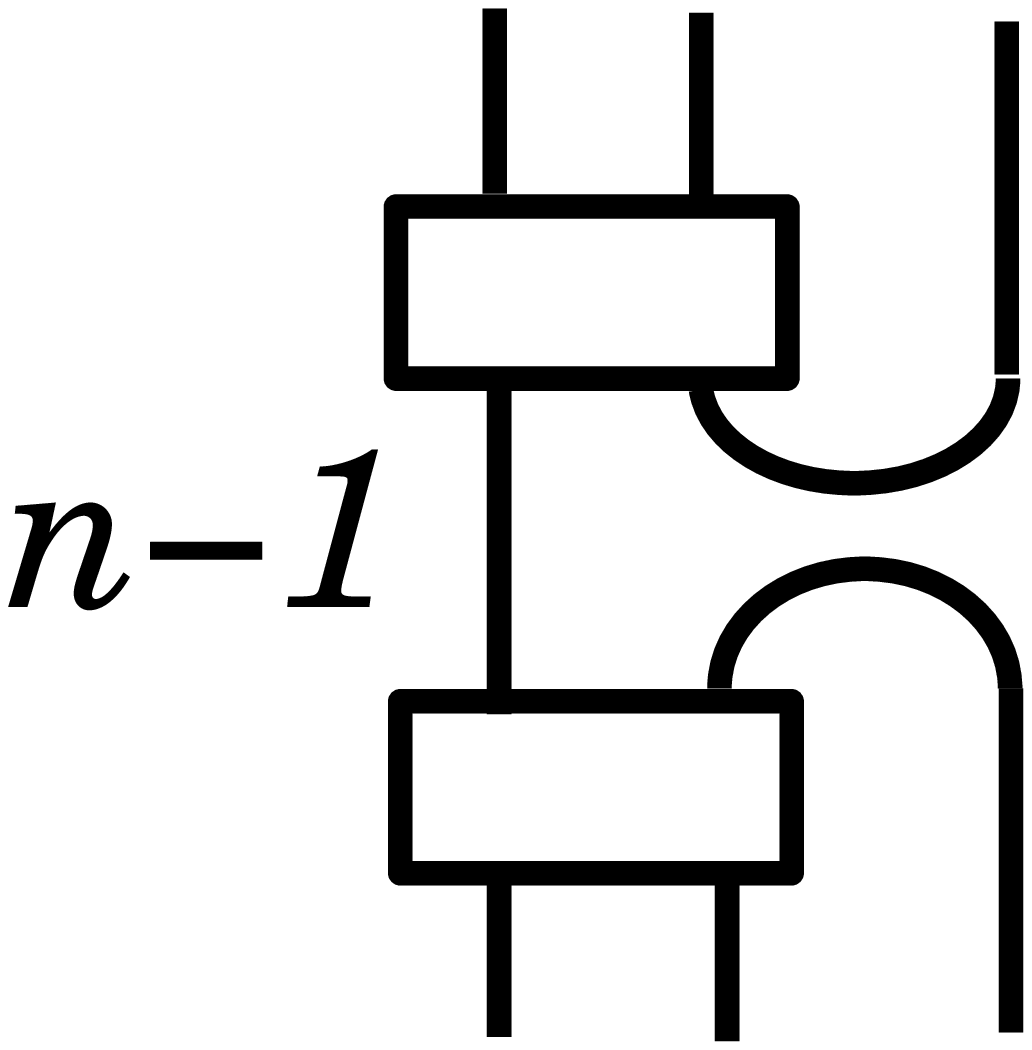}{35}{-15}{-1}{2}.
\label{edgeaddition}
\end{equation}
There is as well an extension of the $\hat{T}^1$ operators, denoted
$\hat{T}^{a}_{q}[\alpha](s)$, where $s$ is a point on the framed loop
$\alpha$.  Their action is defined by,
\f
\langle \Gamma |\hat{T}^{a}_{q}[\alpha](s)=
l_{Pl}^{2}\sum_{I}j_{I}\Delta^{a}[e_{I},\alpha(s)]
\langle e_{I}\#\alpha(s)|
\ff
where the $e_{I}$ are the edges of the spin network, $j_I$ are
the corresponding colors and $\langle e_{I}\#\alpha(s)|$ denotes the
action of
grasping of the ``hand'' of the $T$-operator on the link $e_{I}$. The
action amounts to creating a new four-valent vertex. (Unless
the point of coincidence is already a vertex, in which case the
valence of the vertex is increased by two.)

\begin{figure}
\begin{tabular*}{\textwidth}{c@{\extracolsep{\fill}}c@{\extracolsep{\fill}}
c@{\extracolsep{\fill}}}
\pic{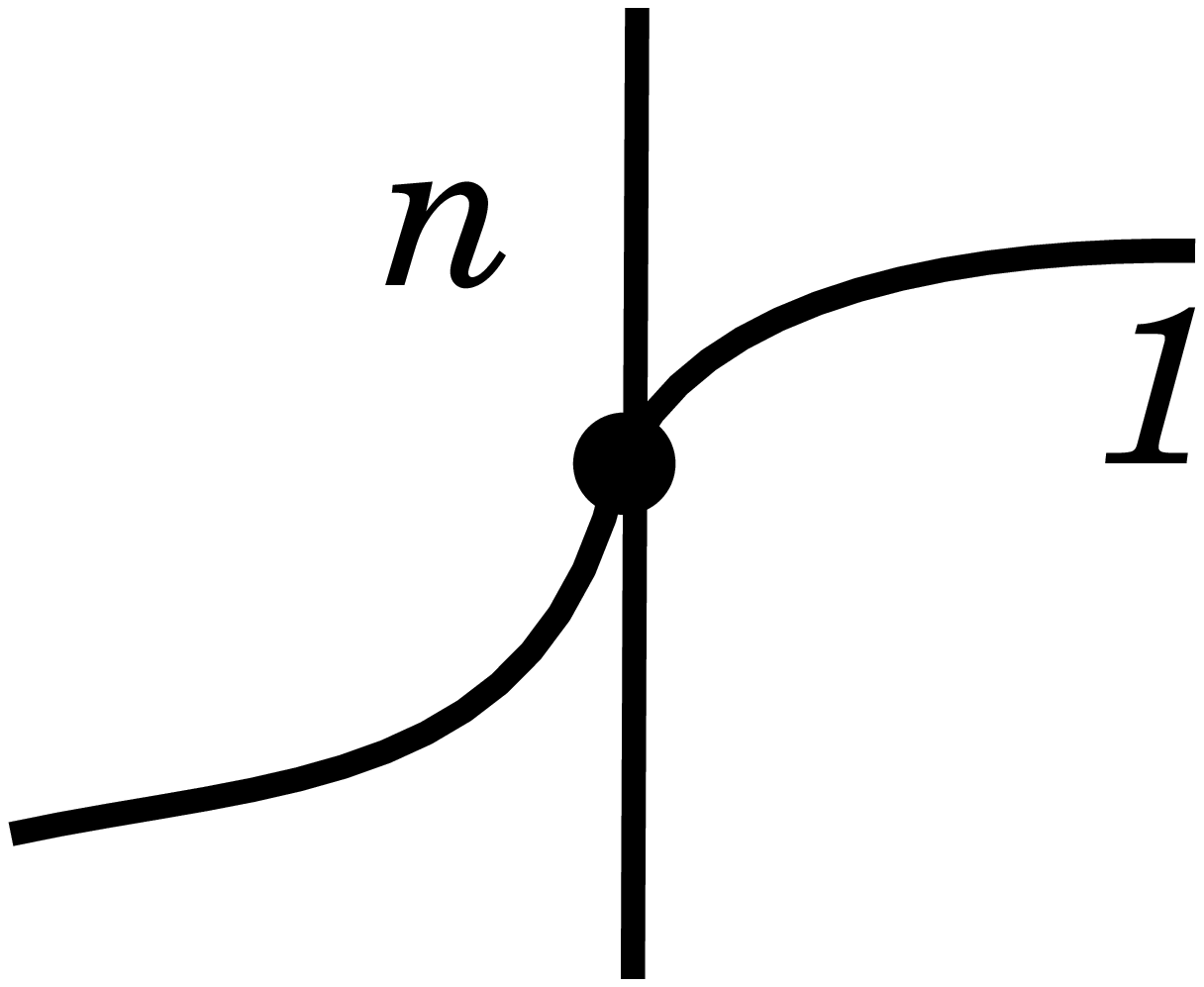}{75}{5}{0}{0} & \pic{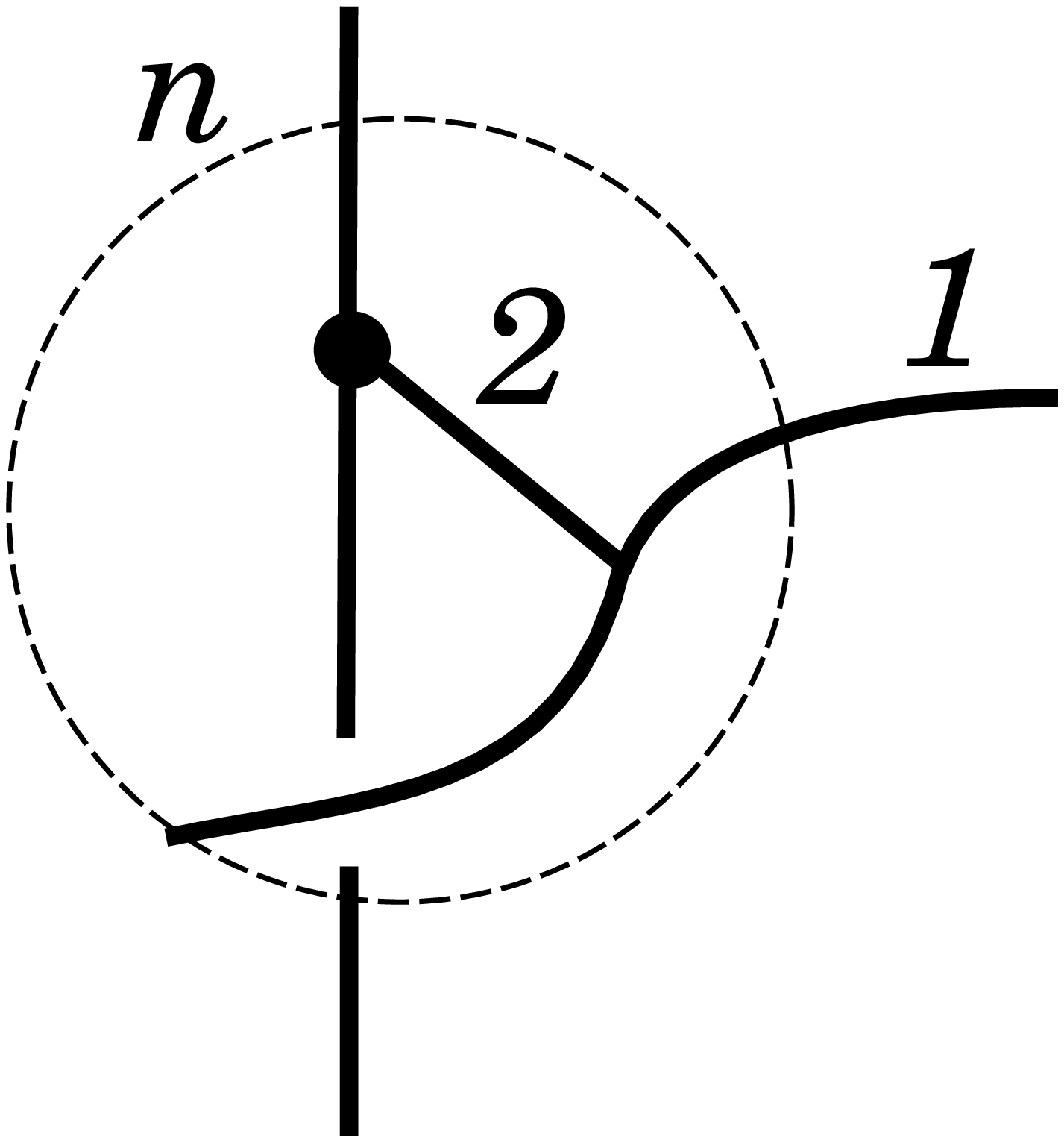}{75}{5}{0}{0} &
\pic{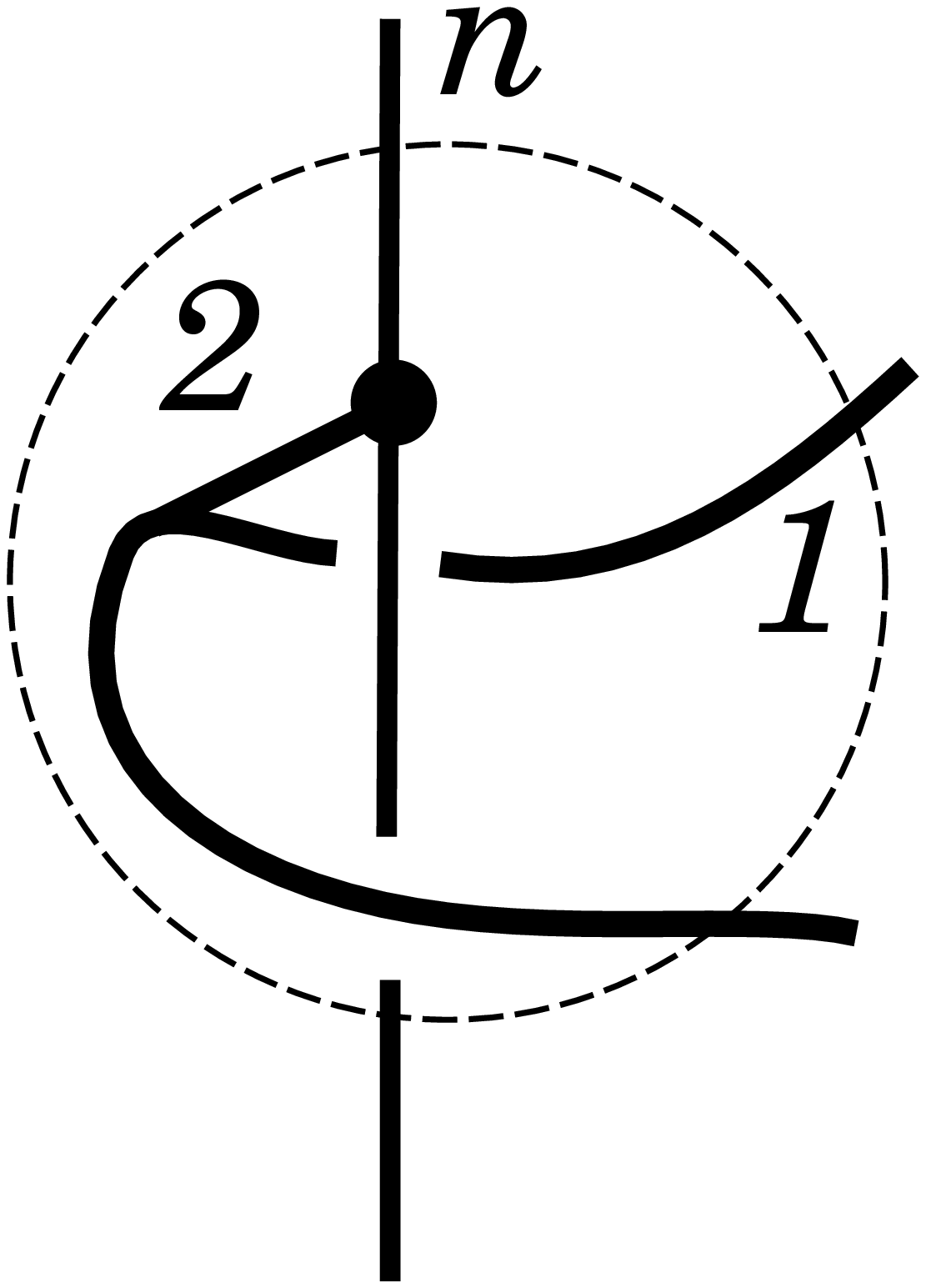}{75}{5}{0}{0} \\
a & b & c \\
\end{tabular*}
\caption{The ambiguity in the action of the $\hat{T}^1$ operator (a.) shown
in (b.) and (c.) corresponding
to a choice of linking.}
\label{amb}
\end{figure}

There is an ambiguity in the definition of the four-valent
vertex corresponding to the action of the ``hand" of the
$\hat{T}^{a}_{q}[\alpha](s)$.  This is illustrated by Fig.(\ref{amb}) which
shows possible ways to decompose the vertex in Fig.(\ref{amb}b and c).
These correspond to additional operator ordering ambiguities which
arise due to the phase factors that can appear in quantum spin networks
from twistings and braidings of edges.
In the construction
of operators through regularization procedures these ambiguities must
be resolved so as to give well defined operators.  As in
the case of operator ordering ambiguities that appear elsewhere
in quantum theory we know of no general prescription for resolving
them.  However, in particular cases natural orderings
arise, as we will see below.

Physically interesting operators in quantum gravity are constructed
by higher loop operators that have more than one site on the loop
at which the action we have just defined takes place.   For
example, the action of a $\hat{T}^{ab}_{q}[\alpha](s,t)$ is given by the
expression:
\f
\langle \Gamma|\hat{T}^{ab}_{q}[\alpha](s,t)=
l_{Pl}^{4}\sum_{I,J}j_{I}j_{J}\Delta^{a}[e_{I},\alpha(s)]
\Delta^{b}[e_{J},\alpha(t)]\langle \Gamma\#_{I}\#_{J}\alpha|
\ff
where $\langle \Gamma\#_{I}\#_{J}\alpha|$ represents the
graphical action of grasping.
Each four-valent vertex defined by the action of a hand
must be defined by decomposing the vertex into a
pair of trivalent vertices, joined with an ``internal"
line of color two as shown on Fig.(\ref{amb}).

The $\hat{T}$ operators with three and more hands are then defined
by extending the definition of the $\hat{T}^2$ so that the action of
each hand is defined according to Fig.(\ref{amb}).  Again, when the
operator is used in the regularization of a particular observable
choices must be made of additional phases coming from the operator
ordering ambiguities associated with the $q$-deformation.

\section{Regularization and recoupling}

Operator products may be dealt with in this formulation of
quantum gravity just as they are in the ordinary loop
representation \cite{ls-review,ham1,volume1}.  A classical diffeomorphism
invariant
observable ${\cal O}$,
is rewritten as a limit, in which one or more
parameters, $\delta$, are taken  to zero, of regulated
observables ${\cal O}_\delta$.   These
are not diffeomorphism invariant as they depend explicitly on
an arbitrary background metric $h^0_{\mu \nu}$ which is used
to define the scale of the point splitting.  This regulated observable
is represented as a quantum operator $\hat{\cal O}_\delta$, which
is constructed from loop operators.  For example, the area,
hamiltonian constraint ${\cal H}$ and
$H=\int_\Sigma \sqrt{-{\cal H}}$ are all represented in terms of
$\hat{T}^2$'s while the volume is represented in terms of
$\hat{T}^3$ as in the ordinary case.

The operators are then defined in terms of the limits
$\delta \rightarrow 0$.  The $c$-number factors must be shown to
assemble into finite factors independent of $h^0_{\mu \nu}$, while
the limits in the dependence of the loop functional must be evaluated
in a topology inherited from the manifold topology, as described in
detail in \cite{ls-review}\footnote{We may note that in the connection
representation it appears to be also necessary to refer to a topology
besides the Hilbert topology in order to define these
limits\cite{ALTMM}.}.
After these limits are taken one has combinatorial expressions
involving loops at a single point.  The basic hypothesis we make
is that the Chern-Simon measure determines the behavior of
the operators in the limit of short distances.  This means that the
Kauffman bracket relations, satisfied by loop factors in the
presence of the Chern-Simon measure, rather than the naive
Mandelstam relations, must be used to evaluate the combinatorics
of loops that are shrunk to a point in a regularization procedure.

We may note that as the theory is not defined by a loop transform
from a measure on the space of connections, such a hypothesis is
necessary to complete the definition of the regularization procedure.

\section{Excerpts from recoupling theory.}

The Kauffman bracket relations are very compactly expressed by
the recoupling theory for the quantum group $SU(2)_q$, which
extends the classical theory of recoupling of angular
momentum\cite{KL}.
The basic relation in this theory expresses the relation between the
different ways in which three angular momenta, say $j_{1}$, $j_{2}$,
and
$j_{3}$ can couple to form a fourth one, $j_4$. The two
possible recouplings are related by the formula:
\begin{equation}
\kd{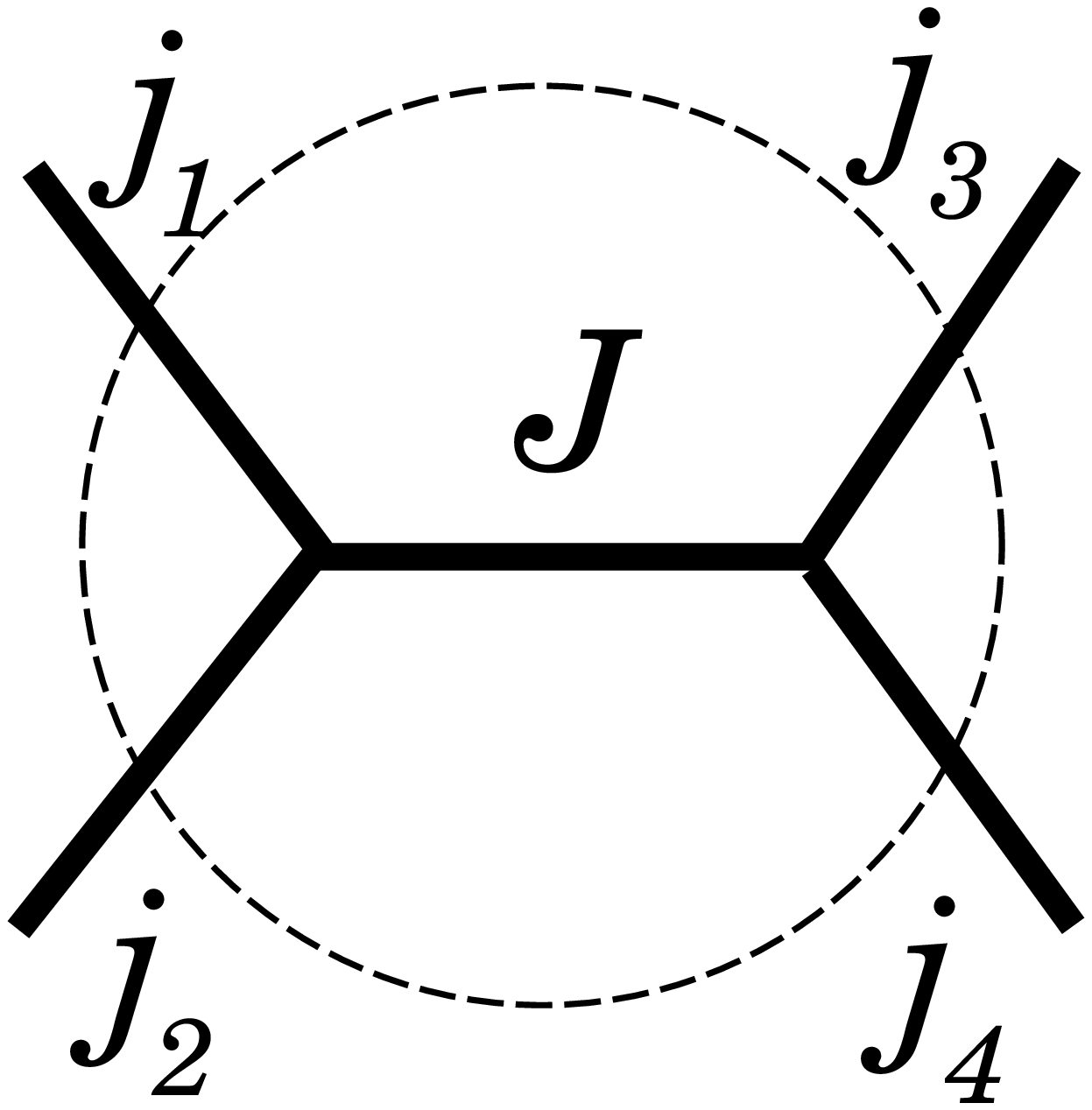} = \sum_I
\left\{ \begin{array}{ccc} j_1 & j_2 & J \\ j_3 & j_4 & I
\end{array} \right\}_{q} \kd{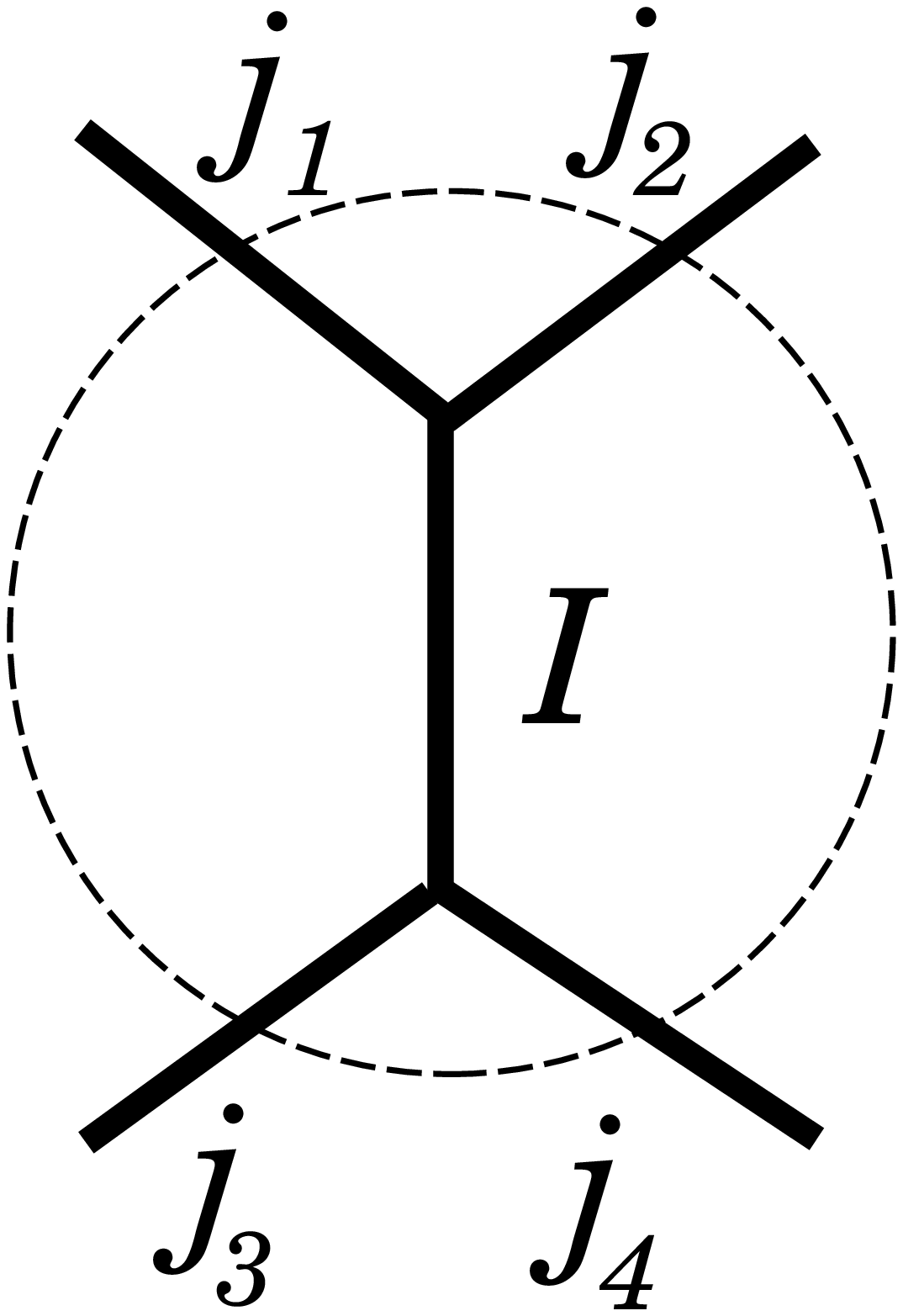}
\label{6j}
 \end{equation}
where on the right hand side is the $q$-6j symbol, as defined in
\cite{KL}.
Closed loops which have
been shrunk to a point may be replaced by their loop value, which is
(for a single loop with zero-self-linking) equal to $-A^{2}-A^{-2}$.  This
extends Penrose's
notion of the evaluation of a closed spin network. The
evaluation of a single unknotted q-spin loop with color $n$ is
\cite{KL}:
\begin{equation}
\pic{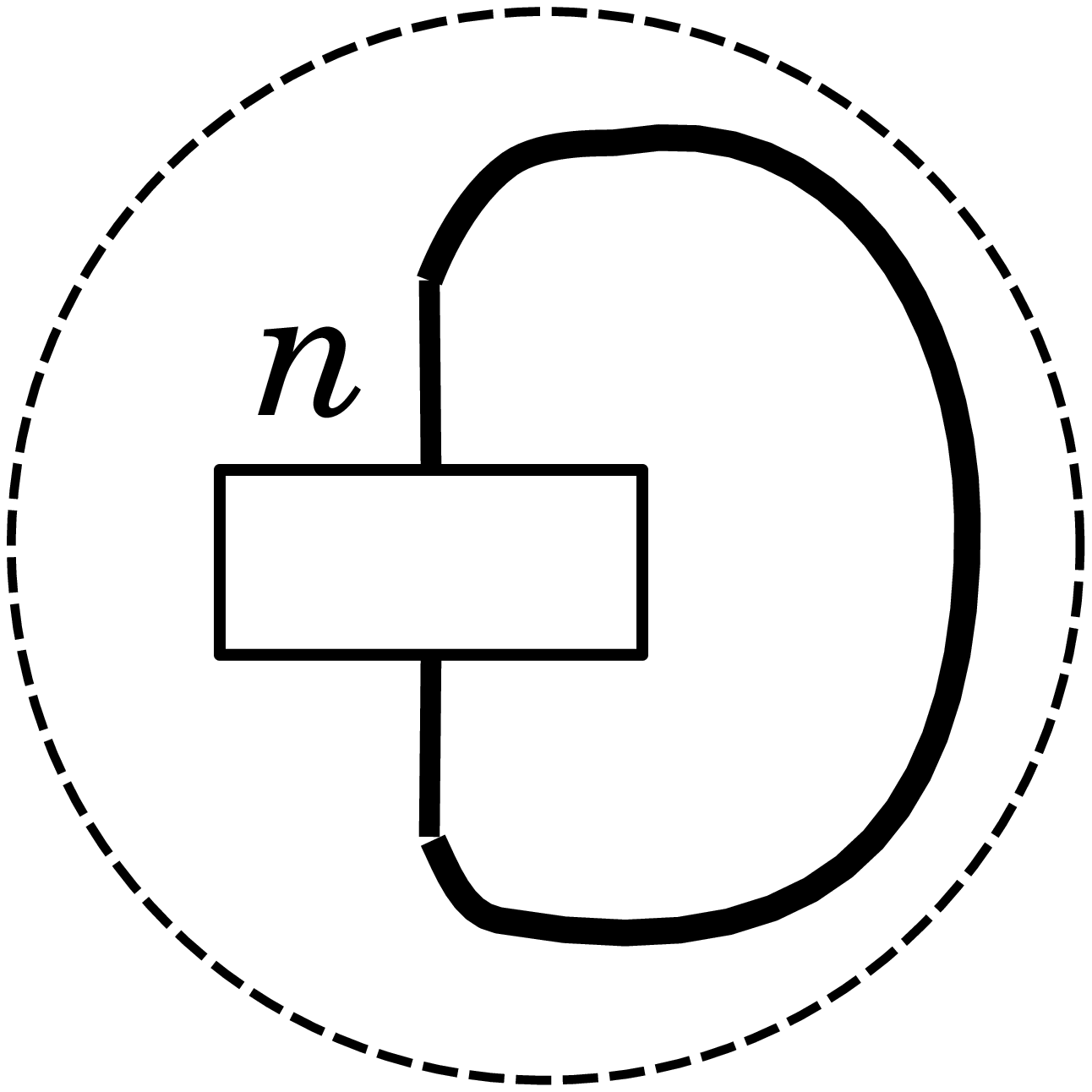}{40}{-15}{0}{0}  = (-1)^n [n+1]
\end{equation}
where $[n+1]$ is a quantum integer.

To see how the given formulae can be used in the evaluation of a q-
spin
net let us consider a ``bubble'' diagram. Upon shrinking of the
``bubble''
this diagram will reduce to a single edge so the evaluation will be
different from zero only if the colors of both ends of the ``bubble''
are
the same. Thus we expect that the ``bubble'' diagram is equal to some
function of the deformation parameter times a single edge. By
closing the free ends of the diagram it is straightforward to
show that
\begin{equation}
\kd{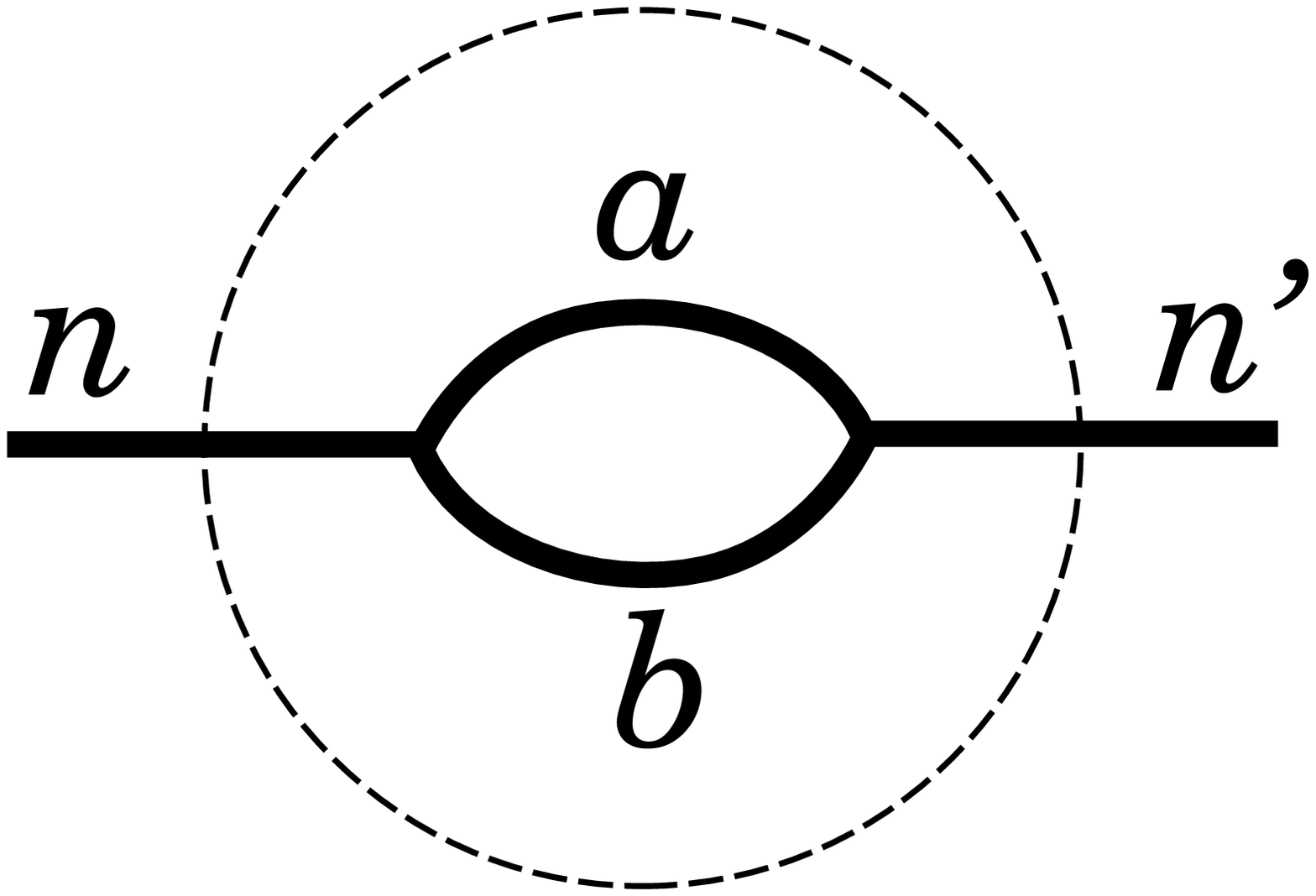} = \delta_{nn'}{ (-1)^{n} \theta(a, b, n) \over [n+1] }
\kd{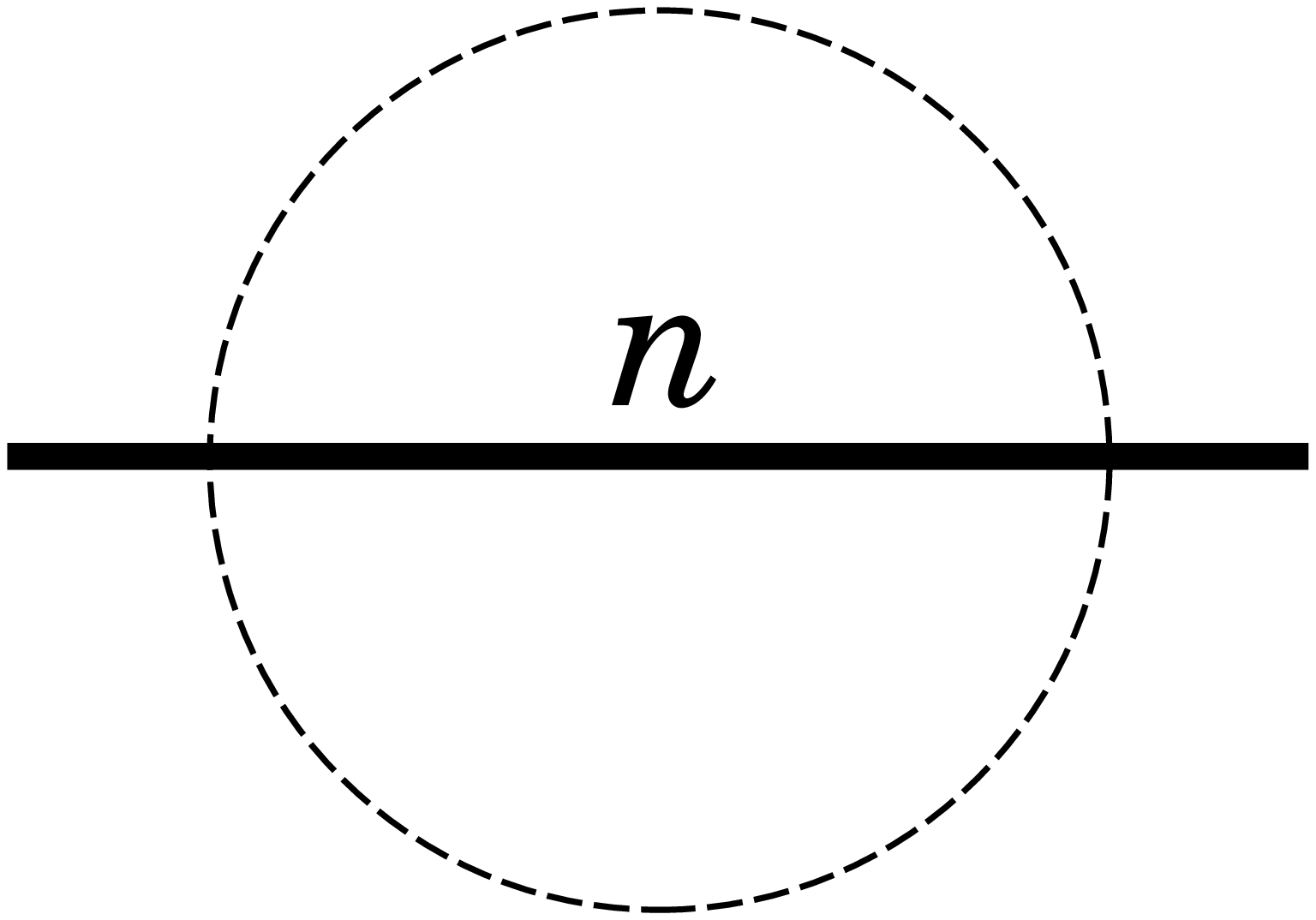}
\end{equation}
in which the function $\theta(a, b, n)$ is given, in general, by
\f\label{theta}
\theta(m,n,l)= \kd{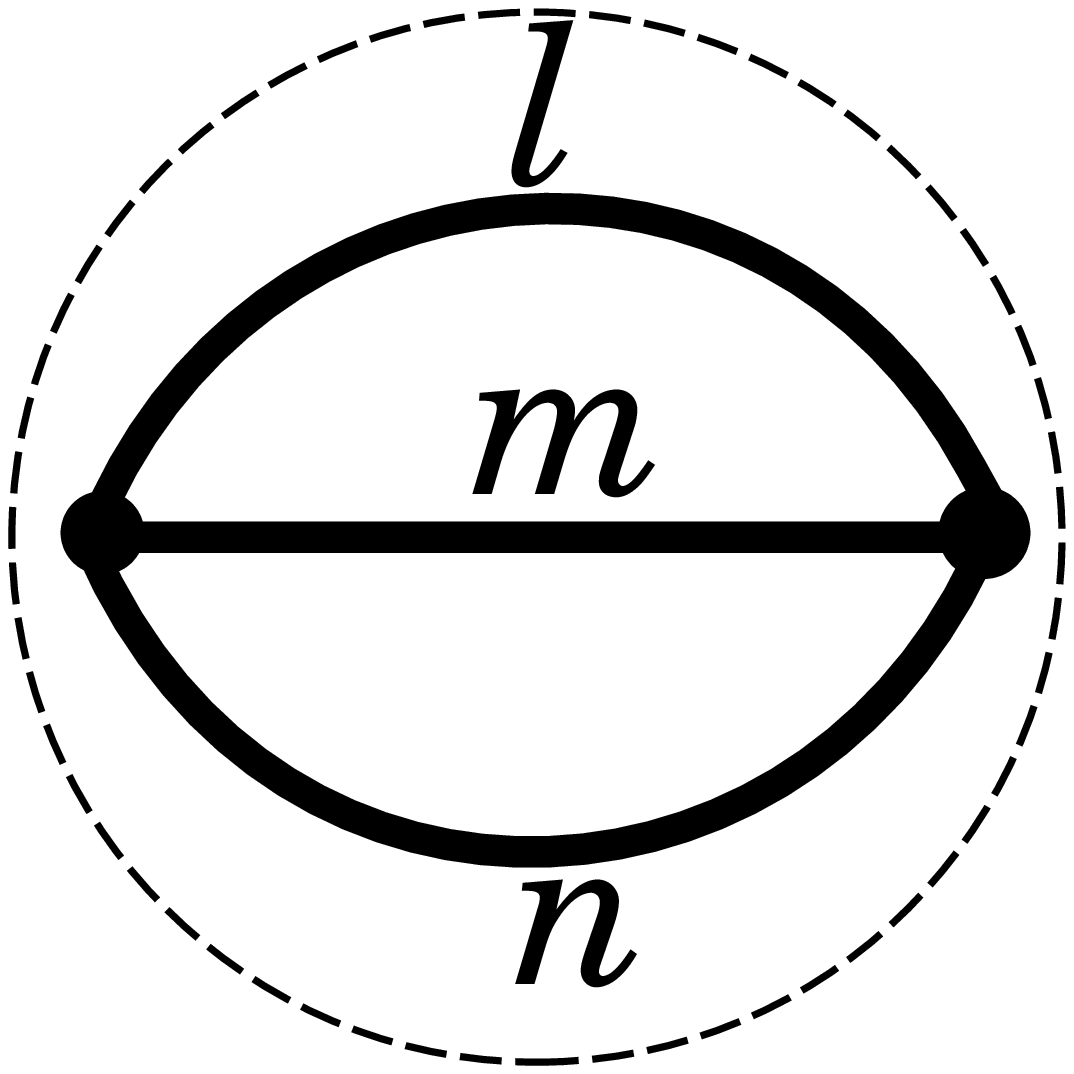} =
(-1)^{(a+b+c)}{[a+b+c+1]![a]![b]![c]! \over [a+b]![b+c]!
[a+c]!}
\ff
where $a+b=m$, $a+c=n$, $b+c=l$ and $[a]!=[a][a-1]...[2][1]$.

A main ingredient in the derivation of the $q$-6j symbol is the tetrahedral
net, which
will be written as $Tet[a,b,e;c,d,f]$. The lengthy expression in terms of
quantum factorials can be found in Kauffman and Lins' book \cite{KL}.

\section{The q-deformed area operator}

We consider
first the regularization of the q-deformation of
the area operator\cite{ls-review,weave,volume1},
which was discussed briefly in \cite{MS}.
We then consider a smooth
2-surface $S$ in the 3-manifold
$\Sigma$ whose area,
$A$ is classically given by
\f
A(S)=\int_{S}d^2\sigma\sqrt{ \left| \tilde{E}^{ai}\tilde{E}^{bi}n_{a}
n_{b} \right| }.
\ff
Using the auxiliary background metric $h^0_{\mu \nu}$, we partition the
surface into small
squares $S_{I}$ of size $L$, so that we have

\f\label{Alimit}
A(S)=\lim_{L\rightarrow 0}\sum_{I}A_{I}=\lim_{L\rightarrow
0}\sum_{I}
\sqrt{A_{I}^{2}}.
\ff
For small surfaces $A_{I}^{2}$ can be approximated by
\f
A_{I}^{2}=\int_{S_{I}}d^{2}\sigma \int_{-\epsilon}^{\epsilon}
{ds   \over 2\epsilon}
\int_{S_{I}}d^{2}\tau  \int_{-\epsilon}^{\epsilon} {dt \over 2\epsilon}
\left| {1 \over 8}
n_{a}(\sigma , s)n_{b}(\tau , t)
T^{ab}_{q}[\alpha](\sigma ,s;\tau , t) \right|.
\ff
Here,  we have a one parameter family of surfaces
$S_I(s)$  displaced
in the background metric normaly to the surface $S$ by a
coordinate distance $s$.  The coordinates on each of these
surfaces are labled by $\sigma$ or $\tau$, and $n_{a}(\sigma , s)$
are the normals at each $\sigma$ and $s$.  (The addition of the
integration in the normal direction, which was not part of the
regularization procedures used before
in \cite{ls-review,weave,volume1,MS} is added here
to resolve the ambiguities in the definition of the
operator.) $T^{ab}_{q}[\alpha](\sigma, s ; \tau ,t)$ is the two-handed
loop variable
based on the q-spin net $\alpha$ passing through the points
$\sigma, s$ and
$\tau, t$.  $\alpha$ can be taken to be a single
$q$-symmetrized edge with its ends at $\sigma,s$ and $\tau,t$.
(In spin network language it is a $2$-line.)
With the above construction we can define a
quantum area operator by
\f\label{Asum}
\hat{A}(S)=\lim_{L\rightarrow 0}\sum_{I}\sqrt{\hat{A}_{I}^{2}}.
\ff
where $\hat{A}_{I}^{2}$ is obtained by replacing $T^{ab}_{q}
[\alpha_{\sigma,s;\tau,t}](\sigma,s;\tau,t)$ by the corresponding
loop operator.  In this loop operator the ambiguities about the
choices of the new four valent vertices and the framing of the
new internal line  are resolved by
choosing $\alpha_{\sigma,s;\tau,t}$ to be the straight untwisted
line between its two end points.

We compute the action of this operator on a state
$\bra{\Gamma}$ whose edges intersect the surface $S$, but are never
tangent to it.
The result of the action is
\begin{eqnarray}
\langle \Gamma|\hat{A}_{I}^{2}&=& l_{Pl}^{4}\int_{S_{I}}d^{2}\sigma
\int_{-\epsilon}^{\epsilon}{ds   \over 2\epsilon}
\int_{S_{I}}d^{2}\tau
\int_{-\epsilon}^{\epsilon}{dt   \over 2\epsilon}
\vert{1 \over 8}n_{a}(\sigma,s )n_{b}(\tau, t)\vert
\nonumber \\
&&\times\sum_{I,J}n_{I}n_{J}\Delta^{a}[e_{I},\alpha(\sigma,s)]
\Delta^{b}[e_{J},
\alpha(\tau,t)]\langle \Gamma\#_{I}\#_{J}\alpha|
\end{eqnarray}
where $n_I$ is the color of the $I$th edge.

Because of the presence of $\Delta^a[e_I,\alpha(\sigma,s)]$ and
$\Delta^{b}[e_{J},\alpha(\tau,t)]$ in the last expression, it is different
from zero only if there is an edge from the $q$-spin net crossing the
$I$th  square.
We consider only the case in which
there is a single
edge of color $n_{I}$ crossing the $S_{I}$.
(The more general case in which there are
nodes of the spin network in the surface can be treated
but for simplicity we do not carry out the computation here.)
The only nonzero terms are those in which two new four valent
vertices are formed
on the edge at $\epsilon$ above and below the plane, to which
the lines $\alpha$ is attached.  Because of the prescription we
have given $\alpha$ runs between the two points without
self-linking and, for sufficiently small $\epsilon$ without
linking any of the edges of the graph.  The next step will be
to take $\epsilon \rightarrow 0$ so that the line
$\alpha$ shrinks and the two new vertices coincide in the
limit.  In this case we will have,
\f
\langle \Gamma\#_{I}\#_{J}\alpha|=c_{q}(n_{I})\langle \Gamma|
\ff
where $c_{q}(n_{I})$ is the result from the grasping which we can
calculate
using the recoupling theory. The sum in Eq. (\ref{Asum}) reduces to a
sum only
over the intersections between the surface $S$ and the q-spin net
$\Gamma$
so
\f
\langle \Gamma|\hat{A}(S)=l_{Pl}^{2}\sum_{I}\sqrt{\vert {1 \over 8}
n_{I}^{2}c_{q}(n_{I})\vert}\langle \Gamma|
\ff
To calculate $c_{q}(n_{I})$ we will note that the graphical action of
grasping of $T^{ab}[\alpha]$ reduces to the
creation of two new trivalent
vertices and thus a ``bubble'' on the edge $n_{I}$.
Upon shrinking of the loop $\alpha$ we get for
the
factor $c_{q}(n_{I})$
\f
c_{q}(n_{I})= {\theta(n_{I},2,n_{I}) \over (-
1)^{n_{I}}[n_{I}+1]}={[n_{I}+2]
\over [2][n_{I}]}
\ff
The last step in the above equation follows from the use of
(\ref{theta}). Thus finally we get
\f
\langle \Gamma|\hat{A(S)}=l_{Pl}^{2}\sum_{I}\sqrt{{1 \over
8}n_{I}^{2}{[n_{I}+2]
\over [2][n_{I}]}}\langle \Gamma|.
\ff
It is a simple exercise for one to show that this result coincides with
the
result obtained in \cite{MS} where a direct calculation was used.

\section{The q-deformed volume operator}

Classically, the volume of a 3-dimensional region $R$
is given by
\f
V=\int_{R}d^{3}x \sqrt{g}.
\ff
We regularize this expression, following a procedure used in the
ordinary
case \cite{volume1}.   We divide the region $R$
into cubes of size
$L$ (using some background metric) so the classical expression for
the
volume becomes
\f
V=\lim_{L \rightarrow 0}\sum_{I}L^{3}\sqrt{\vert {\rm
det}\tilde{E}(x_{I}).
\vert}
\ff
This expression can be expressed in terms of the limit of regulated
observables as
\f
\hat{V}=\lim_{L \rightarrow 0}\sum_{I}{1 \over \sqrt{2^{7}3!}}\sqrt{
{ \hat {\cal W}}_{I}}
\ff
where ${ \hat {\cal W}}_{I}$ is given by the integral
\f
{ \hat {\cal W}}_{I} = \int_{\partial I}d^{2}\sigma \int_{\partial I}d^{2}\tau
\int_{\partial I}d^{2}\rho \left| n_{a}(\sigma)n_{b}(\tau)n_{c}(\rho)
\hat{T}_q^{abc}[\alpha](\sigma,\tau,\rho) \right|.
\ff
We take the framed loop $\alpha$ to be be the triangle formed by
the  straight
(with respect to the background metric) lines between the points
$\sigma$,
$\tau$, and $\rho$.

The action
of the operator ${\hat {\cal W}_{I}}$ obtained in this way on a q-spin
net
will be different from zero only when there is a vertex
in the $I$-th cube. We will consider here  only
q-spin nets with trivalent vertices. The higher valence vertices are
treated in \cite{RB}.   We first compute the action of
${\hat {\cal W}_{I}}$ on a q-spin net graph $\Gamma$
with a trivalent vertex in the
$I$-th box, with edges $m,n,l$  The result is given symbolically by
\f
\bra{\Gamma} {\hat {\cal W}_{I}}  = l_{Pl}^6 mnl \sum_i c_i
\bra{(\Gamma \#\#\# \alpha_{\sigma \tau \rho})_i}
\ff
here $(\Gamma \#\#\#\alpha_{\sigma \tau \rho})_i$ are a finite
set of $q$-spin nets in which each vertex of the triangle
$(\alpha_{\sigma \tau \rho})_i$ is attached to one of the three
edges of the vertex at the points they intersected the box.
There is a sum over possibilities because a choice of ordering must
be made to resolve the ambiguity illustrated in Fig. (\ref{amb})
in the definition of the three new
four-valent vertices in the action of the operator.

The only principle we have to guide this choice is that the operator
should be hermitian as it corresponds to a real quantity.  This
means that we must choose the definitions of the vertices so that
the eigenvalues are real.  The simplest choice that realizes this
is to average over two spin networks
\begin{equation}
{\hat {\cal W}}_{I} \kd{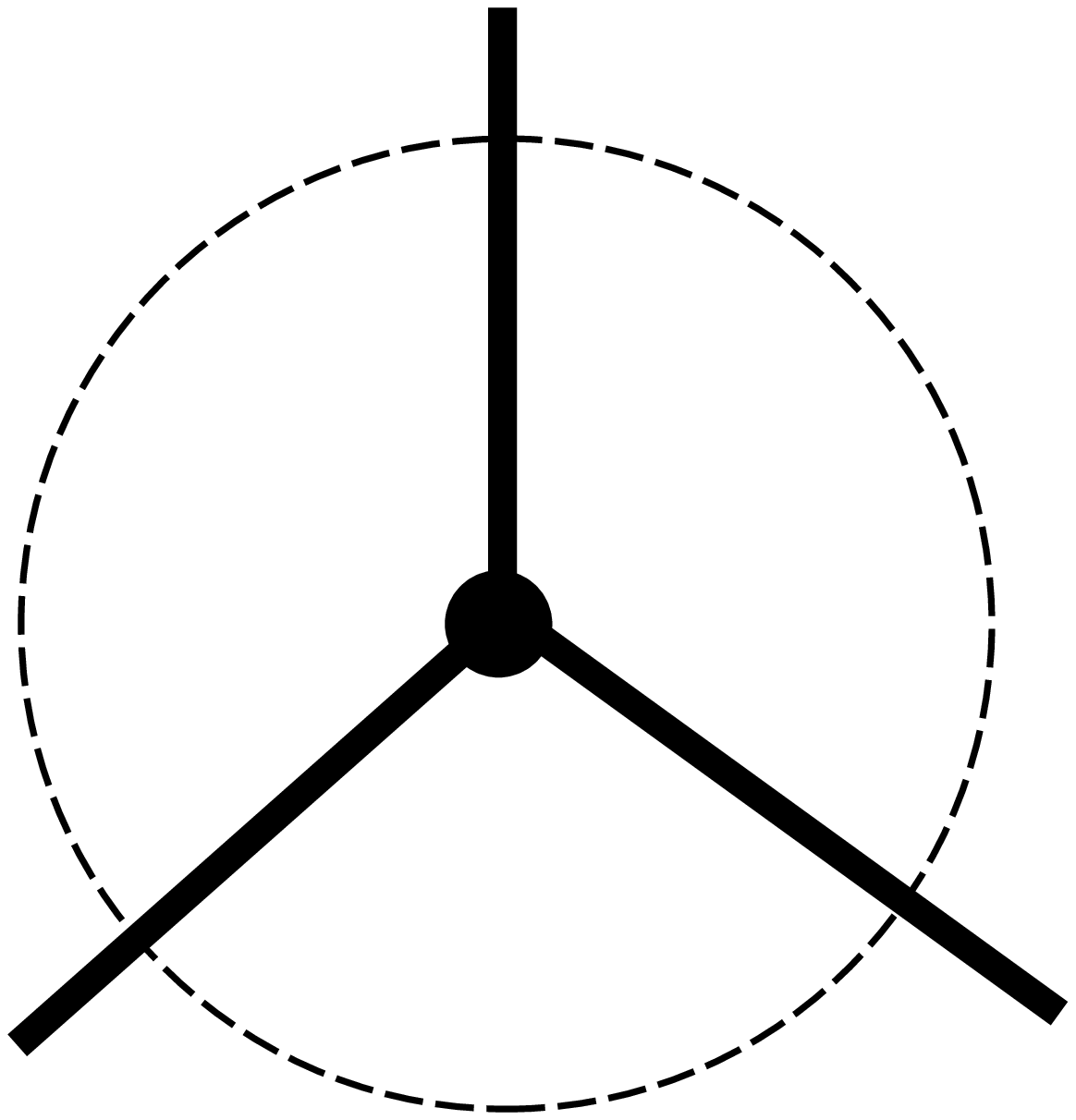} = {1 \over 2} \left[ \kd{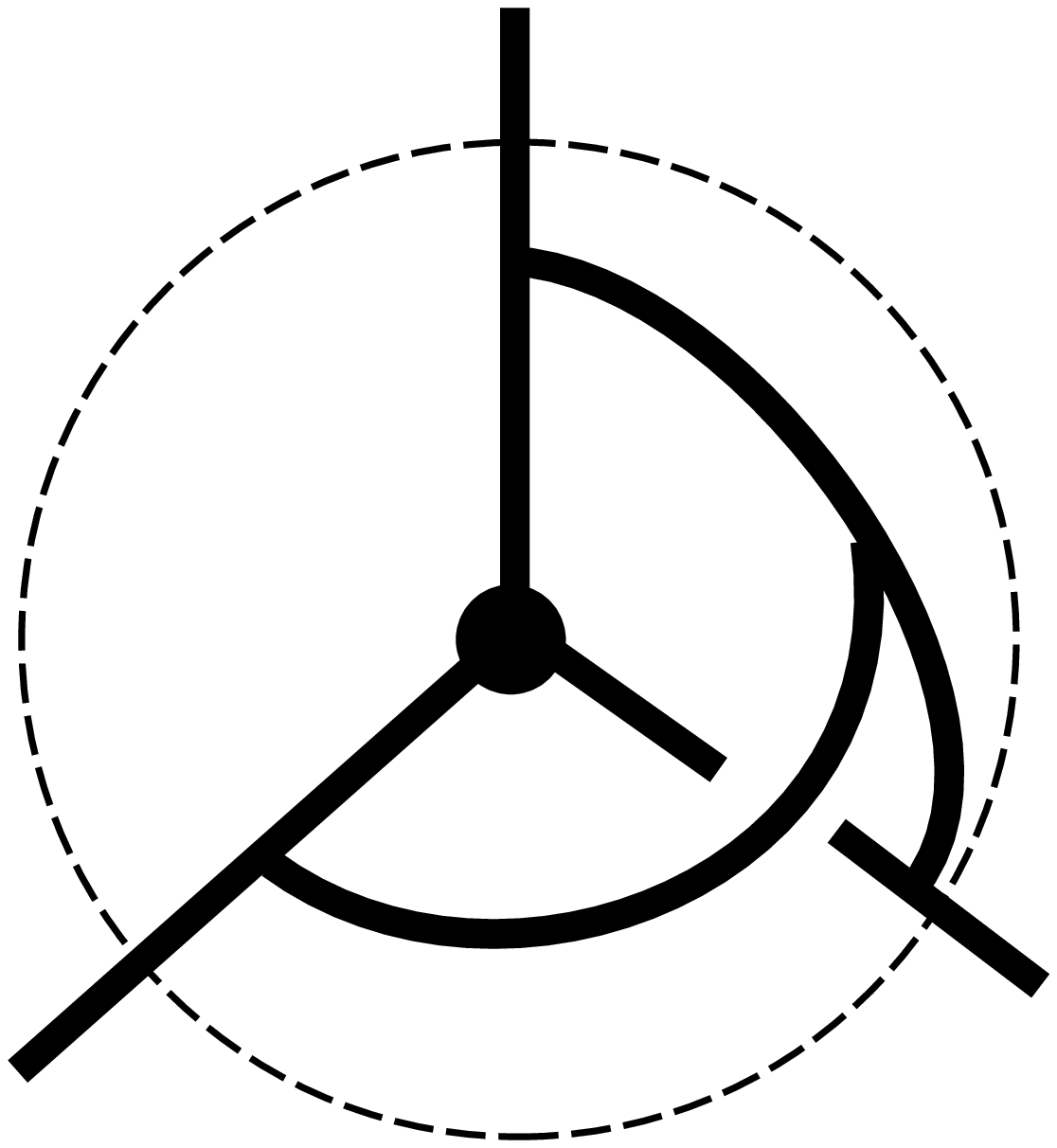} + \kd{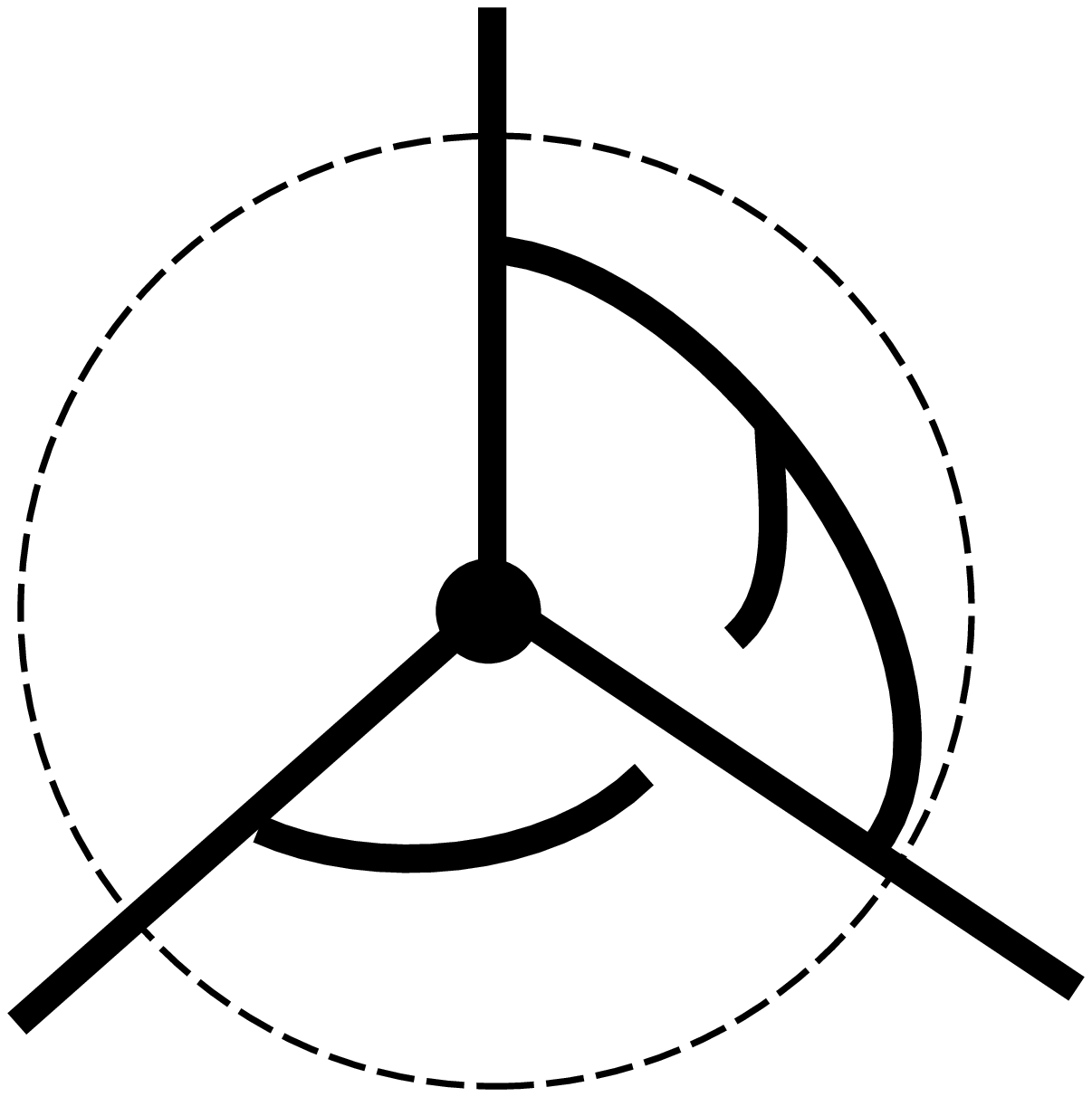}
\right].
\label{vaction}
\end{equation}
where we have shown only the graphical part of the expression.

In each of these the triangle $\alpha_{\sigma \tau \rho}$ has been
deformed smoothly to three edges meeting at a trivalent node,
either to the front or the back of the vertex of $\Gamma$, without
changing the evaluation of the Kauffman bracket.  (This is an
illustration of how loops that are to be shrunk down may be
deformed subject to preserving the Kauffman bracket relations.)

In the limit that $L \rightarrow 0$ we will then have
\f
\langle \Gamma|{\hat V} = {l_{Pl}^{3} \over
4}\sum_{I}\left[m_{I}n_{I}l_{I}{1\over 2}
\left| \sum_{i=1}^2w_{i}(m_{I},n_{I},l_{I}) \right| \right]^{1/2}
\langle \Gamma|
\ff
where the sum $I$ is over the vertices of the graph (which again
are assumed to be all trivalent) and the sum $i$ in each case
is a sum over the two q-spin nets illustrated in Eq. (\ref{vaction}). The
quantities, $w_i(m_{I},n_{I},l_{I})$ are the result of the evaluation
of the
parts of the q-spin net around each vertex containing the additional
edges
coming from the volume operator which shrink to the vertex.
They depend only on
$m_{I}$,
$n_{I}$, and $l_{I}$ which
are the colors of the edges joined at the $I$-th
vertex.

The two diagrams in Eq. (\ref{vaction}) are related to each
other by a parity operation.  But with $q$ at a root of unity, the
action of parity in the evaluation of a Kauffman bracket corresponds
to complex conjugation.  Hence
$w_1(m_{I},n_{I},l_{I})=\overline{w_2}(m_{I},n_{I},l_{I})$, so that the
average is real.

Thus, we have defined a diffeomorphism
invariant prescription for the
action
of the volume operator on q-spin nets having only trivalent vertices.
In
the next section we will compute $w_i(m_{I},n_{I},l_{I})$ using the
recoupling theory.

\section{Eigenvalues of the volume operator for trivalent vertices.}

We can now evaluate the graphs of Eq. (\ref{vaction}) in order to
extract the
eigenvalues of the volume operator \cite{ls-review,volume1,RL}
corresponding to trivalent
vertices of quantum spin nets.
The graphical part of the action can be calculated with the use of the
recoupling theory of the angular momentum \cite{KL}.

We will work out the result for the first term on Eq. (\ref{vaction})
and
will discuss on the differences for the second term. Let us
define $w(m,n,l)$ as a sum, $w(m,n,l) = w_1(m,n,l) + w_2(m,n,l)$,
representing the contributions from the two diagrams in Eq. (\ref{vaction}).
Because the routing of
the loops through the trivalent vertex is unique the trivalent vertex
will then represent an eigenstate of the volume
operator:
\begin{equation}
{\hat {\cal W}}_1 \kd{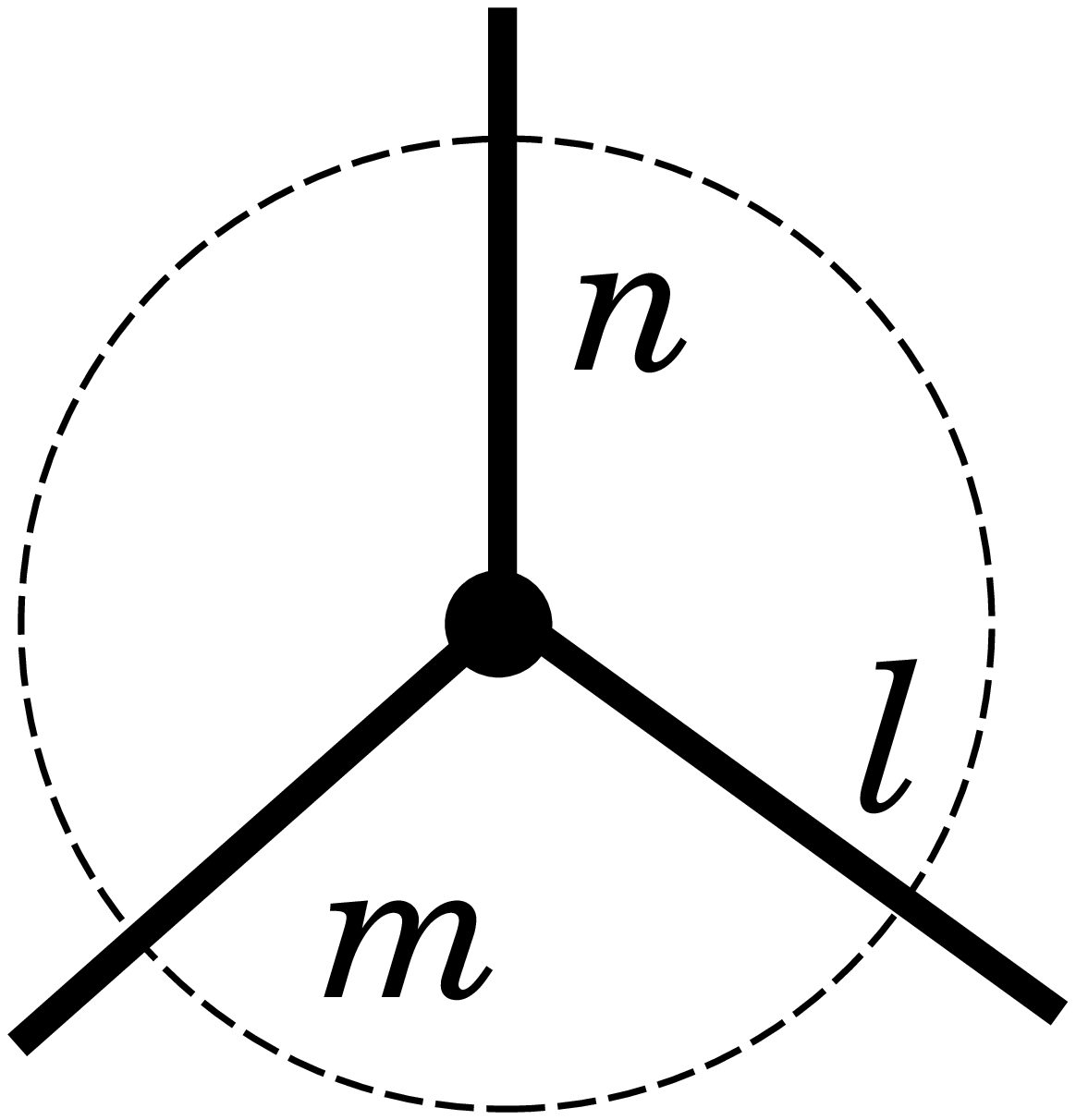} = w_1(m,n,l) \kd{labstv}
\label{wtri}
\end{equation}
where $w_1(m,n,l)$ is the corresponding eigenvalue, which is to be
determined.
The next step for us is to view the vertex cut from the
spin
network and closed to form a new spin network.
By Eq. (\ref{wtri}) it also should be true that:
\begin{equation}
 \kd{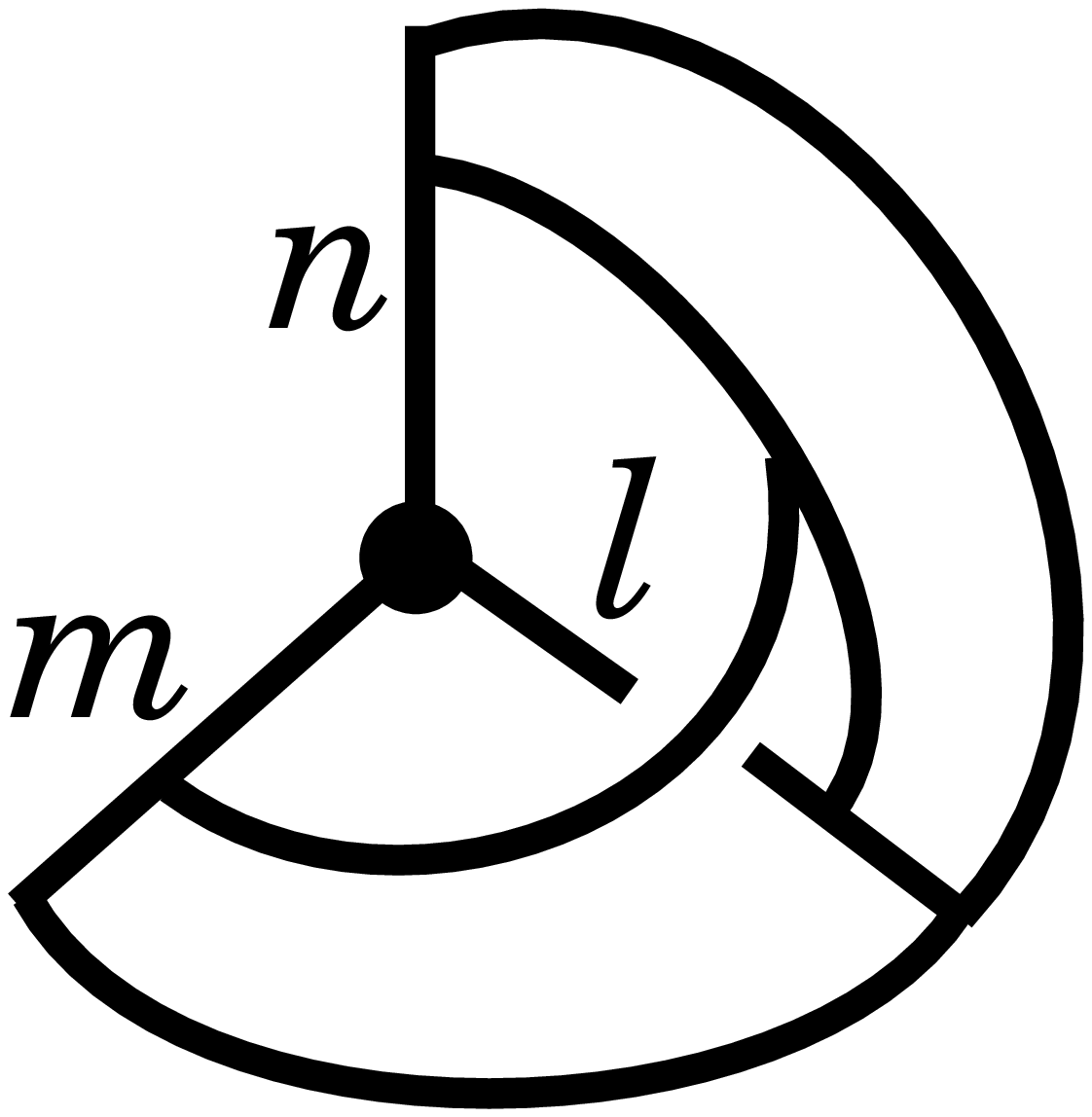} = w_1(m,n,l) \kd{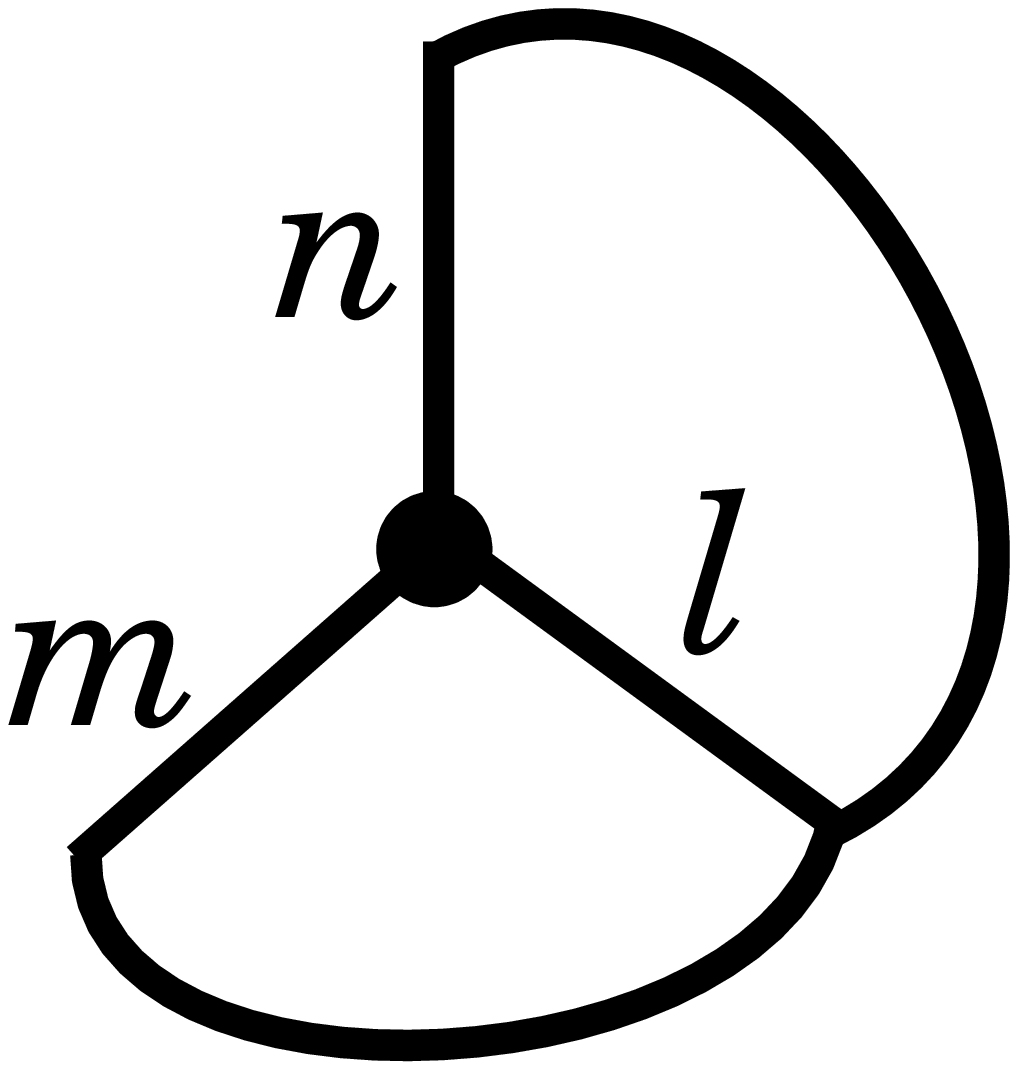}
\end{equation}
(We no longer draw dotted circles around the networks to indicate
that they are being evaluated at a point.)
Thus, in diagrammatic form the $w_1(m,n,l)$ is given by:
\begin{equation}
w_1(l,m,n) =  \left[ \kd{va1closed} \right]  \left[
\kd{thetanet} \right]^{-1}
\end{equation}
The graph in the denominator is the  $\theta$-net. We will evaluate
the
numerator by using the basic formula of recoupling theory. We apply
first
the identity of Eq. (\ref{6j}) to one of the $m$-edges to get:
\begin{equation}
\kd{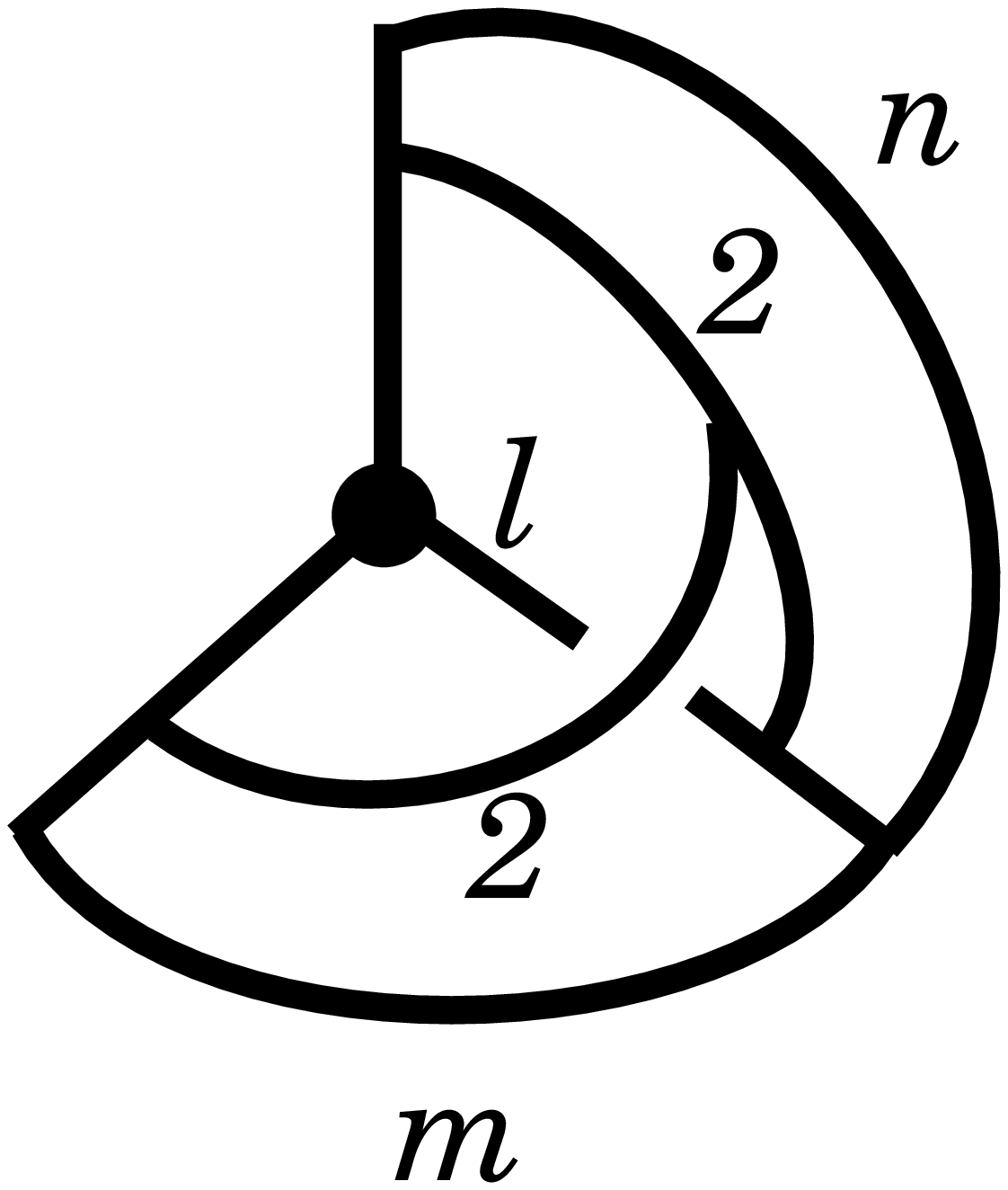} = \sum_j
\left\{ \begin{array}{ccc} 2 & m & j \\ n & l & m \end{array}
\right\} \kd{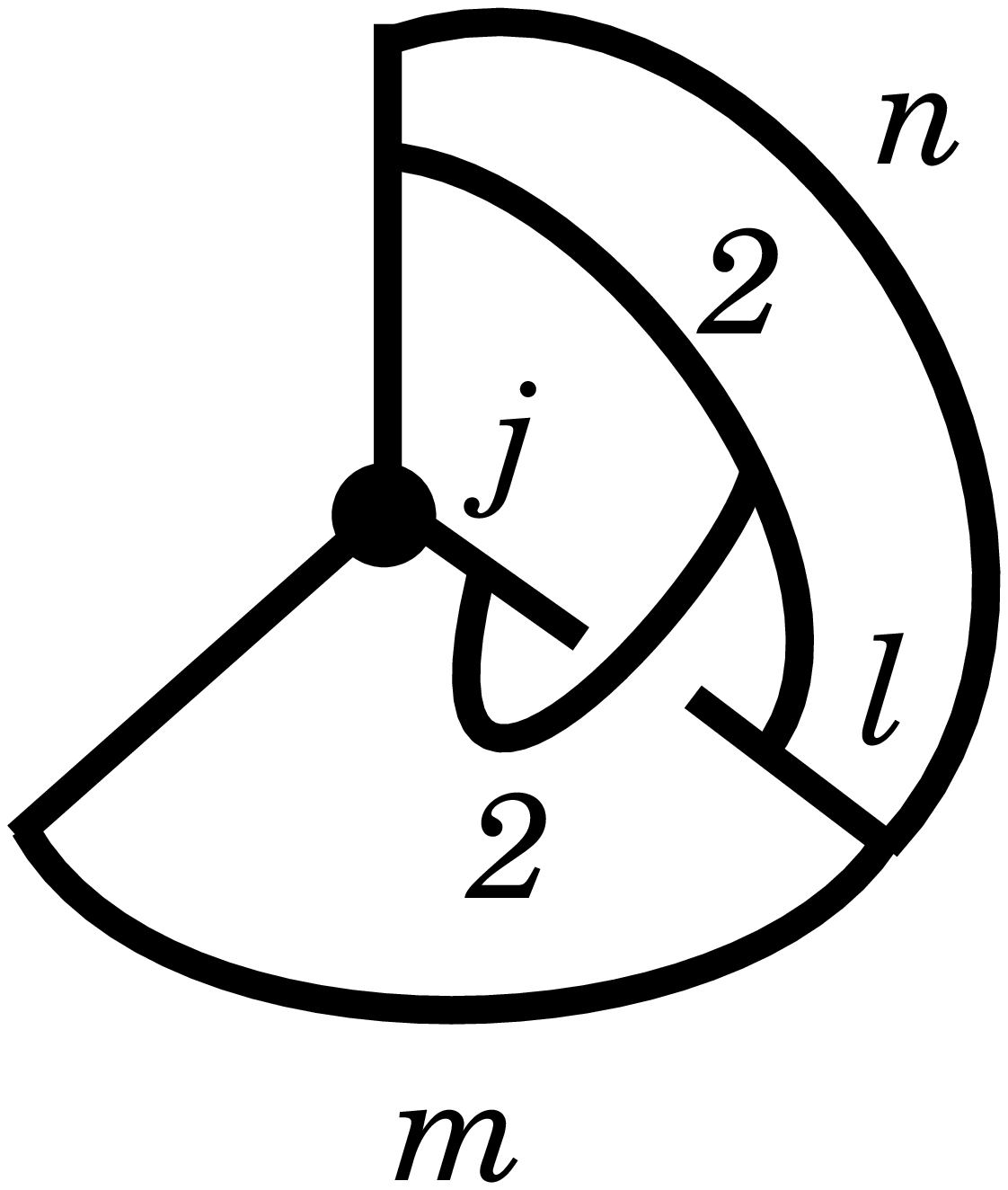}
\end{equation}
Next we use the following identity \cite{KL},
\f
\kd{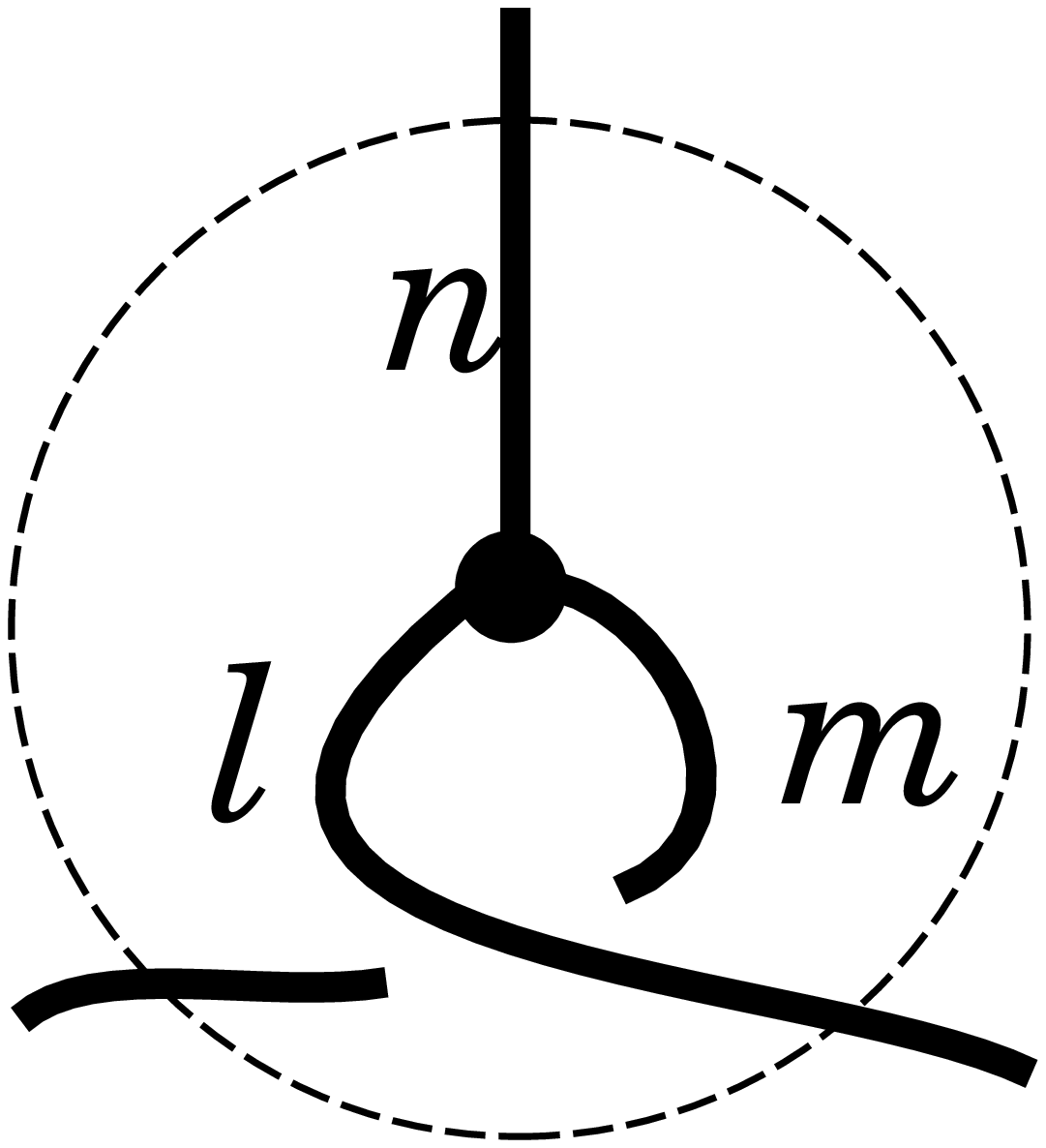} = \lambda^{lm}_n \kd{labstv}
\ff
where $\lambda$ is
\f
 \lambda^{lm}_n= (-1)^{(l+m-n)/2} A^{[l(l+2) +m(m+2) - n(n+2)]/2}.
\ff
We may note that this step is the only place where the difference
between the two terms in Eq. (\ref{vaction})
shows up. At the corresponding step, the second term
in Eq. (\ref{vaction})
will pick up $\lambda^{-1}=\bar{\lambda}$
instead of $\lambda$.
We  then apply the basic recoupling formula (36)
two more times in each term to get finally:
\f\label{main3}
w(l,m,n) =
{1 \over 2}\sum_{j=l-2,l,l+2} (\lambda^{2l}_{j}+
(\lambda^{2l}_{j})^{-1})
\ff
\[{ (-1)^{j}[j+1]{\rm Tet}[2,m,j;n,l,m]
{\rm Tet}[2,2,l;l,j,2]{\rm Tet}[2,n,j;m,l,n] \over \theta(m,n,j)
\theta(m,n,l)\theta(2,l,j)^{2}}.
\]
We may note that the eigenvalue is generally real, as we expected.
Further evaluation is
tedious due to the factorials of quantum integers,
but it is easily done
on a computer, using Mathematica. As an
example, the eigenvalues of the vertex with the lowest admissible
colors
$w(2,1,1)$ is, from this formula or worked out directly,
\f
w(2,1,1)={(1-A^{4})^2 (1+A^8) \over 2 A^4 (1+A^4)^2}.
\ff
We see that this vanishes in the ordinary case when $A = -1$.

There are general
arguments that, in the ordinary case, the volume of trivalent
vertices vanishes \cite{RL,A3}.  Recoupling theory provides
a simple argument for this. First, note that
the general expression for the volume Eq. (\ref{main3})
is invariant under switching
$m$ and $n$; the only effect is to switch the first and third
tetrahedra.  This agrees with the fact that the labeling of the
graphs is arbitrary.  Then, performing a
third Reidemeister move and a twist we have
\begin{equation}
\kd{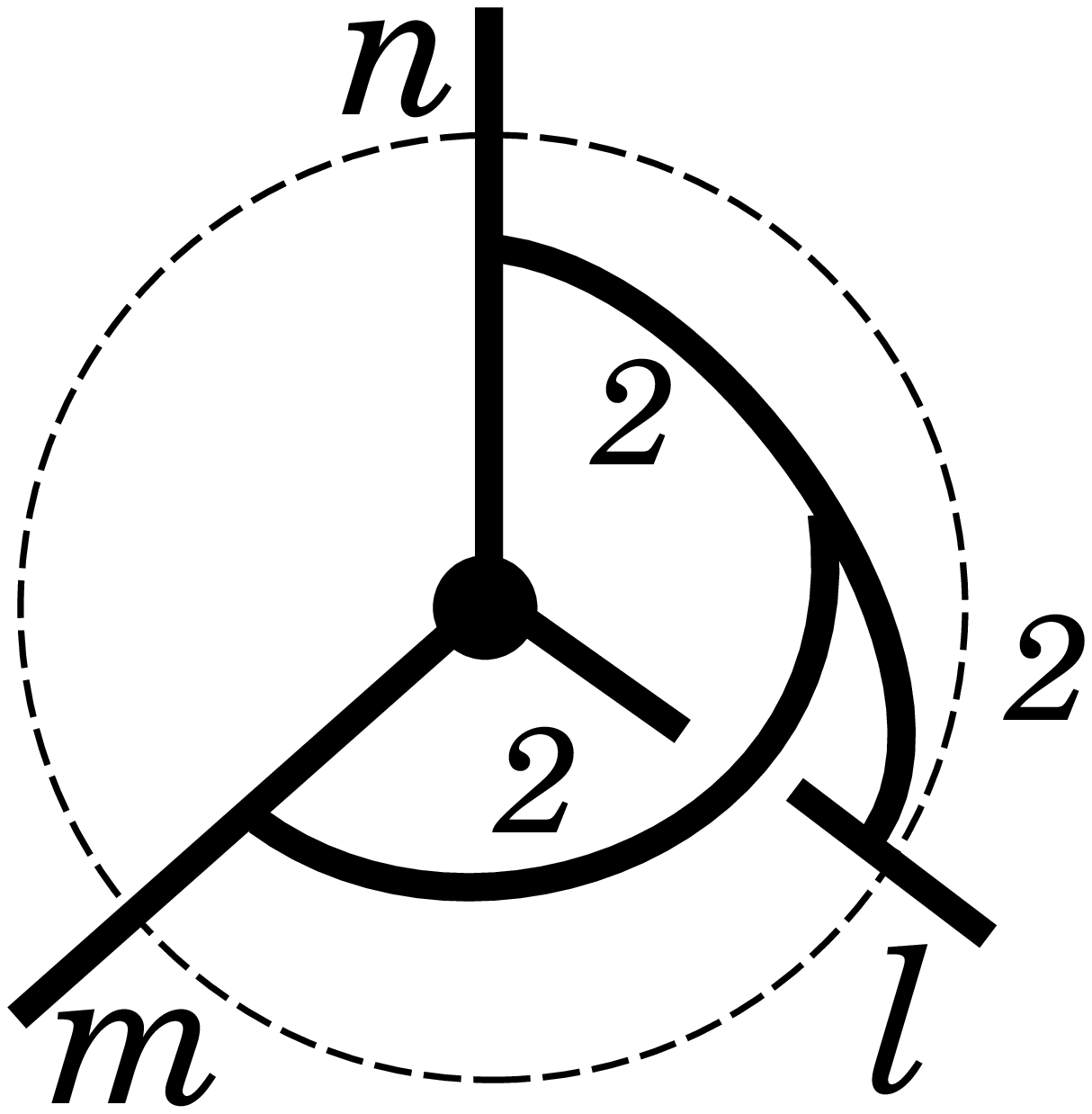} = \kd{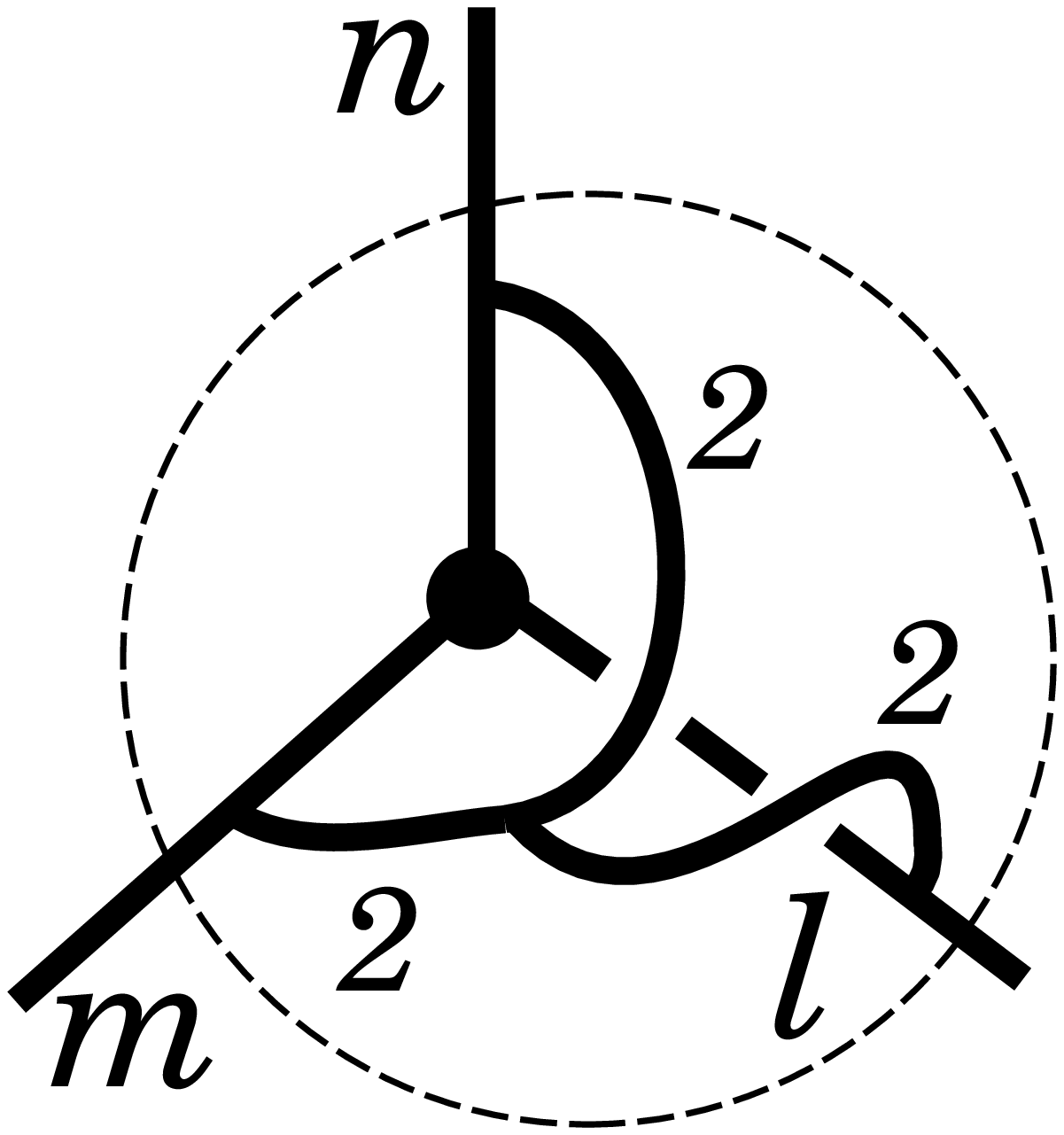} = \lambda_l^{2l}|_{A=-1} \kd{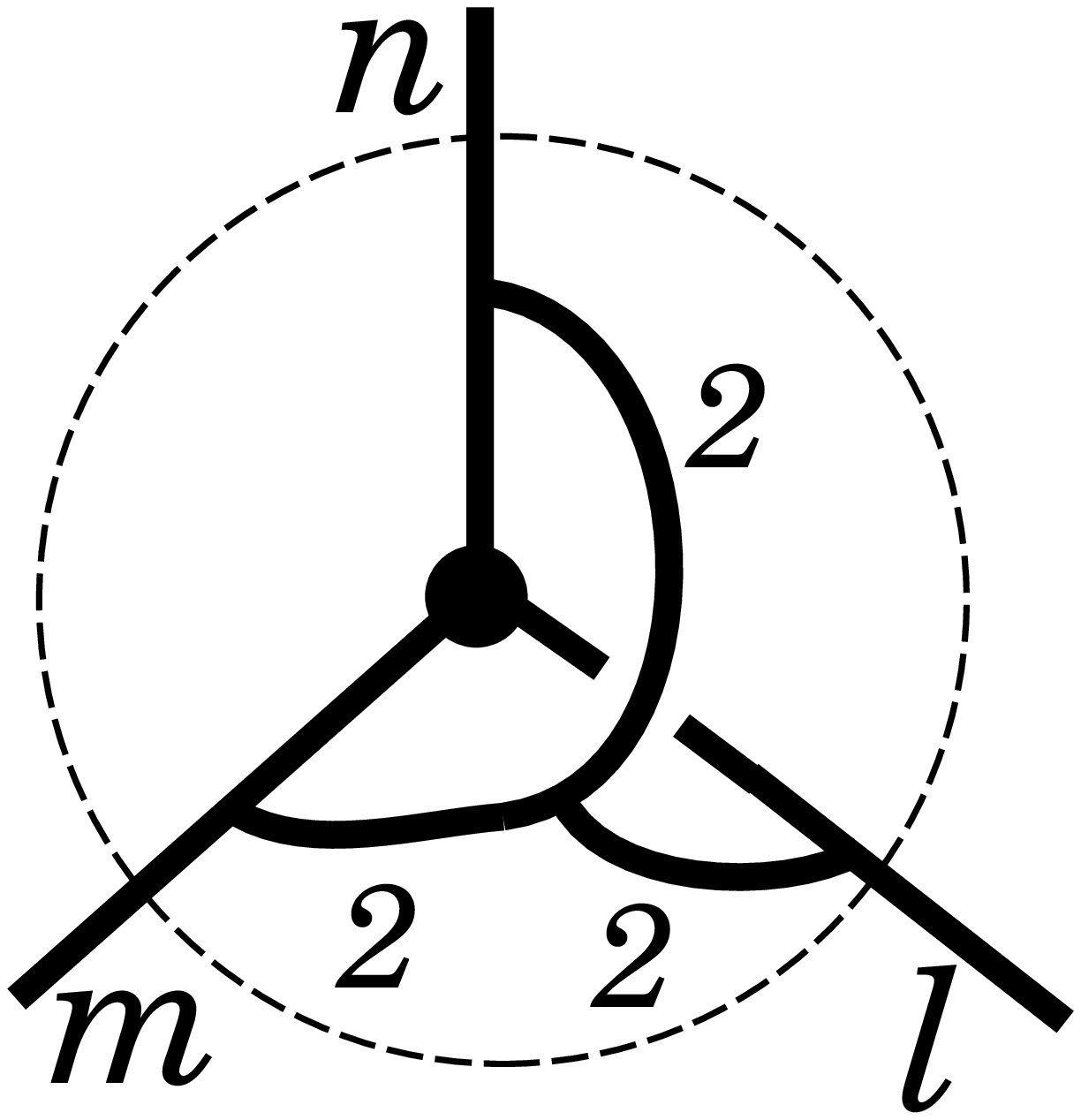}.
\end{equation}
However, $\lambda_l^{2l}|_{A=-1} =-1$, so that, with the invariant
property, the result gives
\begin{equation}
\kd{vv1} = - \kd{vv3} = - \kd{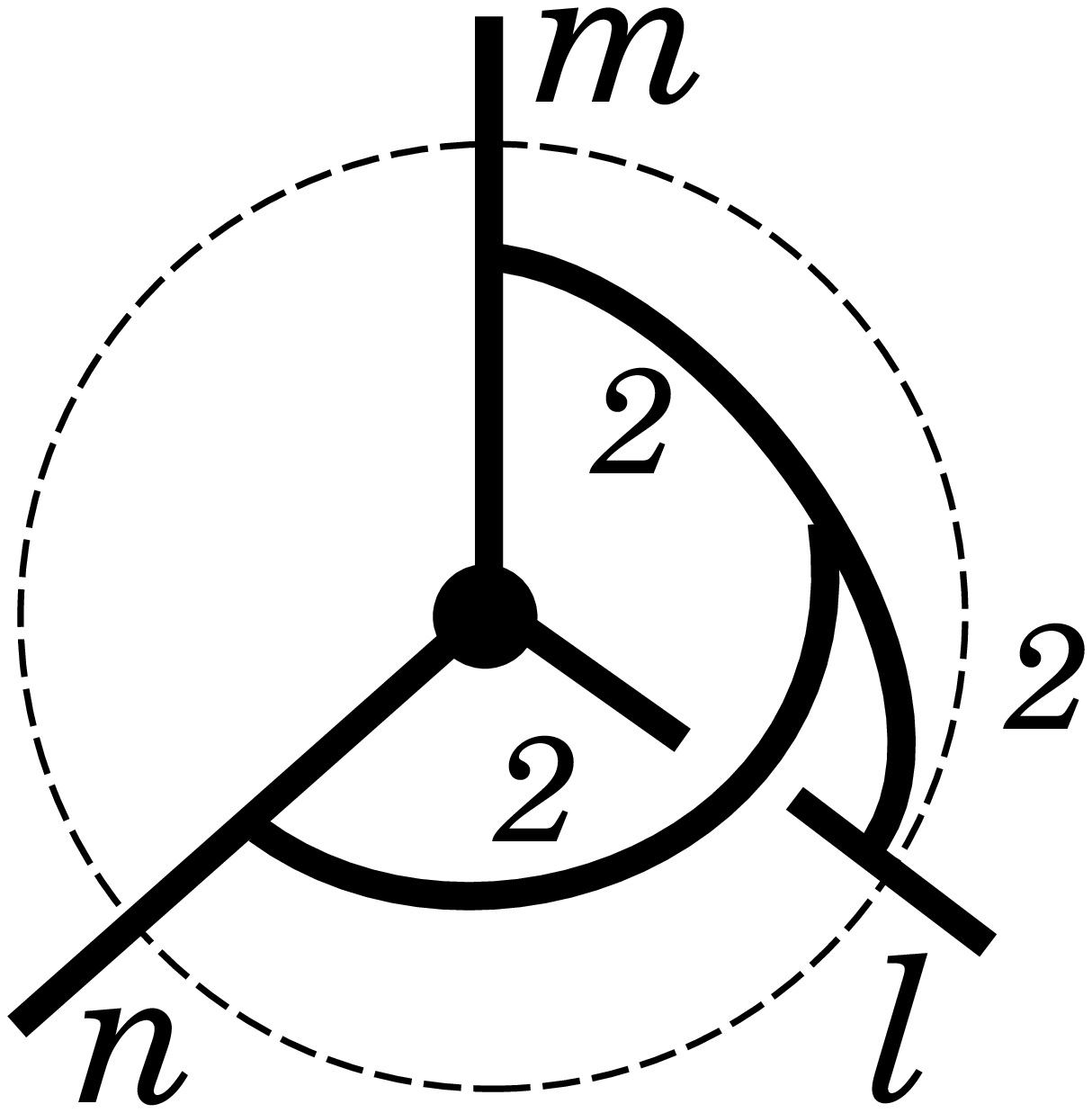} = - \kd{vv1}.
\end{equation}
Thus, an the evaluation is equivalent to the negative of itself; the
volume must vanish.\footnote{The authors thank the participants of
the Warsaw Workshop, especially John Baez, for discussions leading to
this argument.}

\section{Concluding remarks.}

We expect that the results and techniques that we have
described here may be useful for several directions
of further work.

First, the recoupling theory provides an efficient means of
computation that may be readily extended to the computation
of the action of the volume, Hamiltonian constraint and
hamiltonian for general spin networks
of arbitrary valence \cite{BMS}, both in the ordinary and the
$q$-deformed case.  The fact that the degeneracy of the
volume is at least partially
lifted in the $q$-deformed case may make possible the construction
of a variety of interesting operators that involve powers of
the inverse of $det\tilde{E}^{ai}$.  This may make possible
the construction of a strong-coupling expansion for quantum
gravity \cite{BRS},  the evaluation of Thiemann's Wick rotation
operator \cite{thomas-wick,abhay-wick} and the construction
of Hamiltonians corresponding
to interesting gauge choices \cite{lscs}.  For this reason, even if one
is not interested in the hypothesis that the q-deformation
is required to quantize gravity nonperturbatively when the
cosmological constant is non-vanishing, it may still be useful
to regard the quantum deformation as a kind of
diffeomorphism invariant infrared regularization.

Beyond these we may note that the formulation we have defined
here suggests the existence of a class of diffeomorphism invariant
quantum field theories whose basis of states is spanned by the
diffeomorphism invariant classes of embeddings of the spin networks
of an arbitrary quantum group.  These are a large class of theories
that most generally are defined in terms of modular tensor
categories \cite{TV,louis2d3d,louisdavid,louis-int,baezdolan}.
It is natural to extend the
definitions of area and volume operators to these theories. By doing
so we can interpret each of them as diffeomorphism invariant
quantum theories which unify
spatial geometry with other degrees of freedom.  Dynamics may
be postulated for such theories combinatorially,
by generalizing the action of the hamiltonian constraint of quantum
gravity on quantum spin networks \cite{BRS} to these cases.
It will then be sufficient to discover the
connection to general relativity only in the limit in which
the volume of space becomes large.
The possibility of recovering general relativity from such a limit
of a discrete theory is suggested also by the recent result of
Jacobson; which requires only that there be a relationship between
area and information content \cite{ted-ees}.  However, this relationship may be
preserved in these extensions, given the results
of \cite{linking,lsbekenstein}.
Finally,
given that the language of such a theory is closely connected to
that of the minimal conformal field theories \cite{moreseiberg},
it is tempting to
speculate that such a formulation might provide a link between
non-perturbative formulations of quantum gravity and string theory.

\section*{Acknowledgments}

We wish to thank Abhay Ashtekar, John Baez, Louis Crane,
Louis Kauffman,
Viqar Husain, Erik Martinez, Don
Neville,  Mike Reisenberger, Carlo Rovelli and Morten Weis for
discussions and
encouragement during the course of this work.  This work was
supported in part by NSF grant number PHY-93-96246 to the Pennsylvania
State University.

Also one of us (R.B.) is gratefull to the CGPG for the hospitality.

\end{document}